\documentclass[a4paper,11pt]{article}
\usepackage{a4wide,amsmath,amssymb,bbm}
\usepackage[english]{babel}

\def \XX {{\rm X}}
\def \p  {\phi}\def \del {\partial}

\def \ed {\end{document}}

\makeatletter

\@addtoreset{equation}{section} \makeatother

\newcommand \rf [1] {(\ref{#1})}

\begin{document}

\hfill{Imperial-TP-LW-2016-02}

\vspace{10pt}

\begin{center}
{\LARGE{\bf Kappa-symmetry of  superstring sigma model \\
\vspace{10pt}
and generalized 10d supergravity equations}}

\vspace{30pt}

A.A. Tseytlin\footnote{Also at Lebedev Institute, Moscow. tseytlin@imperial.ac.uk }
 \  \  and\ \    L. Wulff\footnote{l.wulff@imperial.ac.uk   }
\vspace{5pt}

{\it\small Blackett Laboratory, Imperial College, London SW7 2AZ, U.K.}\\

\vspace{30pt}

{\bf Abstract}
\end{center}
\noindent
We determine the constraints imposed on the 10d  target superspace geometry by the requirement of classical 
kappa-symmetry  of  the Green-Schwarz superstring. In the type I case we find that the background must satisfy a  
 generalization  of type I supergravity equations. These equations   depend on an  arbitrary vector $X_a$   and imply the one-loop   scale invariance of  the  GS  sigma model.   In the special case when $X_a$ is the gradient of a scalar   $\p$ (dilaton)  one recovers  the standard type I  equations  equivalent to the 2d Weyl invariance conditions of the superstring  sigma model. In the type II case we find a  generalized version of the 10d supergravity equations the   bosonic part of which  was  introduced in arXiv:1511.05795. These equations depend on  two vectors $\XX_a$   and $K_a$  subject to 1st  order differential  relations  (with  the equations in the 
NS-NS  sector depending  only on  the combination  $X_a = \XX_a + K_a$).  In the special  case of  $K_a=0$ 
one finds that $\XX_a=\del_a \phi$   and thus  obtains   the standard type II  supergravity  equations.
New generalized solutions  are found    if $K_a$   is chosen to be a  Killing vector 
(and thus they exist only if the metric admits  an isometry).  Non-trivial solutions of the  generalized equations   describe $K$-isometric  backgrounds that can be mapped  by T-duality  to type II supergravity solutions 
with dilaton containing  a linear isometry-breaking term.  Examples of   such backgrounds  
appeared recently  in the context of    integrable  $\eta$-deformations of $AdS_n \times S^n$    sigma models. 
The    classical kappa-symmetry  thus   does not, in general, imply the 2d Weyl invariance conditions 
for  the GS  sigma model  (equivalent to type II supergravity equations)  
but only weaker  scale invariance type  conditions.

\

\vspace{100pt}

\pagebreak 
\tableofcontents
\setcounter{page}{1}
\def \ci {\cite}
\def \te {\textstyle} \def \ha {{\te {1\ov2}}} \def \ov {\over}
\def \foot {\footnote}\def \la{\label}
\def \be {\begin{equation}}
\def \ee {\end {equation}}

\def \bi {\bibitem}

\section{Introduction  and summary}

The purpose of this paper is to determine precisely which constraints the presence of the 
kappa-symmetry of the Green-Schwarz (GS)  superstring places on the target space (super) geometry. In an influential 1985 paper \cite{Witten:1985nt} Witten showed that the equations of motion of   10d  super Yang-Mills theory  can be expressed as integrability along light-like lines
 and showed that this condition  is closely connected to kappa-symmetry of the superparticle
in the super Yang-Mills  background. 
 He also suggested that there should be a similar connection between the kappa-symmetry of the 
 GS superstring and the supergravity equations of motion. 
 Shortly thereafter,  the type II GS  superstring action in a general supergravity background was written down in \cite{Grisaru:1985fv} and  it was  shown  that the standard on-shell
 superspace constraints of type IIB supergravity \cite{Howe:1983sra} are sufficient for the string to be  kappa-symmetric.  It was    conjectured that these  constraints  are  also necessary for   the  kappa-symmetry.

In \cite{Shapiro:1986yy} it  was   found  that for the type I superstring the kappa-symmetry implies the basic (i.e. dimension 0 and --$\ha$)  superspace constraints on the torsion and 3-form $H=dB$ superfields\foot{There is also a constraint for the  Yang-Mills sector which we will ignore here.
 In our notation $a,b,...=0,1,2,...,9$ are bosonic tangent space indices,  and $\alpha,\beta,...=1,...,16$ are 10d Majorana-Weyl spinor indices.}
\begin{equation}
T_{\alpha\beta}{}^a=-i\gamma^a_{\alpha\beta}\,,\qquad \qquad H_{\alpha\beta\gamma}=0\,,\qquad\qquad H_{a\alpha\beta}=-i(\gamma_a)_{\alpha\beta}\,.
\label{1.1}
\end{equation}
Ref.\cite{Shapiro:1986yy}  also argued that these constraints are  enough to make the target space geometry  a solution of type I supergravity.\foot{More recently   \cite{Berkovits:2001ue} it was shown that (classical) BRST invariance of the pure spinor superstring \cite{Berkovits:2000fe} (which may  be  viewed as an 
  analog of kappa-symmetry in this formulation)  implies the basic 
  type I and type II  constraints.     It  was  argued 
   that these constraints are enough to put the corresponding supergravity background completely on-shell (see,  however,  note added in section \ref{sec:conclusion}).
   }

While in  the  11d    case  the condition of kappa-symmetry of the supermembrane action \cite{Bergshoeff:1987cm}
leads to a constraint on the torsion   which   implies that  the  background   should satisfy the standard 11d  supergravity equations of motion  \cite{Howe:1997he}, 
 here we will  find  (in  disagreement   with the earlier  conjectures/claims) 
that this  is not so  in the 10d superstring case: 
the 10d 
 supergravity equations are sufficient  but not necessary   for kappa-symmetry.

In the case of the type I GS superstring  where  the  kappa-symmetry implies the basic constraints (\ref{1.1}),  
we shall show, by  solving completely the Bianchi identities for the torsion and the 3-form,
 that these constraints actually lead to  a weaker set of equations than those of type I supergravity. These equations are similar to the 
 conditions for 1-loop scale invariance of the GS sigma model \cite{Arutyunov:2015mqj}, 
  which are, in general,  weaker  than the  Weyl invariance
  conditions  required  to  define   a consistent superstring theory. 
  This is not totally surprising  as the  condition of classical kappa-symmetry does not take into account the dilaton term $\int d^2 \xi \sqrt g R^{(2)} \phi(x)$  required  to make  the quantum 
  2d stress tensor traceless (see \ci{ca}  and discussion in
   \cite{Arutyunov:2015mqj}).\foot{Some discussions of one-loop 
  quantum corrections in GS   sigma model (in the  heterotic string  case) appeared in \ci{het}.}
  

Indeed,  the problem in 10d  compared to the 11d  case is the presence of the dilaton. The dimension $\ha$  component  of the  torsion   is   expressed in terms of a spinor (``dilatino") superfield $\chi_\alpha$. If one requires that $\chi_\alpha$ is expressed in terms of a scalar superfield $\phi$ (the dilaton) as 
\begin{equation}\la{1.2}
\chi_\alpha=\nabla_\alpha\phi
\end{equation}
then  the Bianchi identities for the torsion  imply 
the standard type I supergravity equations   \cite{Nilsson:1981bn}. 
However,  if  this extra assumption \rf{1.2} (not required for  kappa-symmetry) 
  is dropped, 
 the basic constraints \rf{1.1}  imply only the equations for a
  ``partially off-shell"    generalization of the type I supergravity equations.
The solution of the constraints   and Bianchi identities then depends on an 
arbitrary vector $X_a$ (that replaces the dilaton gradient)\foot{As $X_a$ is subject to a 
constraint on its divergence  we get 8  additional  bosonic fields  compared to 
 the standard type I theory. These  are  matched by an extra 8 
  fermionic  components  present due to  the fact that the dilatino $\chi$ is now off-shell 
  whereas in the standard type I supergravity it satisfies a Dirac equation.}
  and the bosonic equations of motion take the form (here 
  fermionic fields are set to zero)
\begin{align}
\te R_{ab}+2\nabla_{(a}X_{b)} -\frac14H_{acd}H_b{}^{cd}=&0\,,\la{1.3}\\
\te \nabla^cH_{abc}-2X^cH_{abc} -4\nabla_{[a}X_{b]}=&0\,,\la{1.4}\\
\te \nabla^aX_a-2X^aX_a+\frac{1}{12}H^{abc}H_{abc}=&0\,.\la{1.5}
\end{align}
If one  restricts to   the special  case of $X_a=\del_a\phi$,
 these equations reduce to the standard type I supergravity equations of motion
 (or  string effective equations in the NS-NS  sector).
 The  generalized equations \rf{1.3},\rf{1.4}
  coincide with the 1-loop scale invariance conditions of a  bosonic sigma model $L=(G-B)_{mn} \del_+ x^m \del_- x^n$ provided the reparametrization and $B$-field gauge freedom vectors 
  are  chosen to be equal.\foot{In the notation of
  \cite{Arutyunov:2015mqj} this means $Y_a=X_a$.
  This  identification  is a consequence of the underlying supersymmetry  of the 
  equations  leading to  \rf{1.3}--\rf{1.5}. It should  come out  automatically if the scale  invariance of the GS string is studied in the  manifestly   supersymmetric (superspace) form.}
  The conclusion is  that the condition of classical kappa-symmetry is  essentially 
   equivalent  to   the  one-loop scale invariance  condition  for the type I  GS sigma model.
  Only the  stronger condition of 2d Weyl invariance (eqs. \rf{1.3}--\rf{1.5} with 
  $X_a =\del_a \phi$)
  is equivalent to  the standard  type I supergravity equations of motion.

Performing a similar analysis in  the case of the type IIB GS superstring  we will find that the 
kappa-symmetry  implies  the  direct generalization 
of the basic constraints   \rf{1.1} on the torsion and 3-form\foot{Here $i,j,k=1,2$ label the two MW spinors of type IIB superspace and $(\sigma^r)_{ij}$  ($r=1,2,3$) are  Pauli  matrices. The gamma-matrices $\gamma^a_{\alpha\beta}$ and $\gamma_a^{\alpha\beta}$ are $16\times16$ symmetric `Weyl blocks' of 10d Dirac matrices satisfying
 
$\gamma^a_{\alpha\beta}(\gamma^b)^{\beta\gamma}+\gamma^b_{\alpha\beta}(\gamma^a)^{\beta\gamma}=2\eta^{ab}\delta_\alpha^\gamma\,,$ 
see \cite{Wulff:2013kga} for more details on our notation.}
\begin{equation}
T_{\alpha i\,  \beta j }{}^a=-i\delta_{ij} 
\gamma^a_{\alpha\beta}\,,\qquad \qquad H_{\alpha i\, \beta j\, \gamma k}=0\,,\qquad\qquad H_{a\, \alpha i\, \beta j }=-i \sigma^3_{ij} (\gamma_a)_{\alpha\beta}\,.
\label{1.6}
\end{equation}
When the Bianchi identities are solved we will conclude  that  these constraints 
lead  again  not to the type IIB supergravity equations   but to a weaker set of generalized type  IIB
equations involving, instead of the  dilaton scalar,  two vectors $\XX_a$ and $K_a$. 
The  corresponding bosonic equations  may be written as (here as in \rf{1.2}--\rf{1.5} the fermionic
component  fields are set to zero)
\begin{align}
&\te R_{ab}+2\nabla_{(a} X_{b)}-\frac14H_{acd}H_b{}^{cd}+\frac{1}{128}\mathrm{Tr}(\mathcal S\gamma_a\mathcal S\gamma_b)=0\,, \la{17}\\
&\te \nabla^cH_{abc}-2{ X}^cH_{abc}-4\nabla_{[a}X_{b]} -\frac{1}{64}\mathrm{Tr}(\mathcal S\gamma_a\mathcal S\gamma_b\sigma^3)=0\,, \la{18}\\
&\te \nabla^a X_a-2{ X}^a X_a +\frac{1}{12}H^{abc}H_{abc}-\frac{1}{256}\mathrm{Tr}(\mathcal S\gamma^a\mathcal S\gamma_a)=0\,, \la{19}\\
&\te \gamma^a\nabla_a\mathcal S
-\gamma^a\mathcal S\,\mathrm  (\XX_a - \sigma^3 K_a) 
+\big(\frac18 \gamma^a\sigma^3\mathcal S\gamma^{bc}\,
+\frac{1}{24}\gamma^{abc}\sigma^3\mathcal S\,\big)H_{abc}
=0\,. \la{20}
\end{align}
They generalize the type I  equations   \rf{1.3}--\rf{1.5} to the presence of 
   the analog of the  RR  field strength    bispinor   $\mathcal S = (\mathcal S^{\alpha i\, \beta j})$  (which   includes  the factor of $e^\p$  in the standard type IIB case 
   \cite{Tseytlin:1996hs,Wulff:2013kga})
\begin{equation}\la{200}\te 
\mathcal S=-i\sigma^2\gamma^a\mathcal F_a-\frac{1}{3!}\sigma^1\gamma^{abc}\mathcal F_{abc}-\frac{1}{2\cdot 5!}i\sigma^2\gamma^{abcde}\mathcal F_{abcde}\,.
\end{equation}
Combining \rf{17} and \rf{19} we  get  the following 
   generalized ``central charge" equation 
\begin{equation}\la{233}\te 
\bar \beta ^X\equiv R-\frac{1}{12}H_{abc}H^{abc}+4\nabla^a X_a-4{ X}^a X_a=0\,.
\end{equation}
As was shown in  \cite{Arutyunov:2015mqj}, the relation $\del_a \bar \beta ^X=0$
follows, in fact, from eqs.\rf{17},\rf{18},\rf{20}   so that the  ``dilaton equation" \rf{19}
is not indepedent.

In the above equations \rf{17}--\rf{233} 
   \begin{align}
 X_a \equiv  \XX_a + K_a \ , \la{21}\end{align}
and the vectors $\XX_a$ and $K_a$  are subject to  
\begin{align}
& \nabla_{(a}K_{b)}=0 \la{222} \ , \qquad \qquad  {\mathrm X}^aK_a=0 \ ,  \\
&\qquad 2\nabla_{[a}\mathrm X_{b]}+K^cH_{abc}=0 \,.\la{22}
\end{align}
Thus    $K_a$ satisfies  the Killing vector equation, while  eq.\rf{22}  
 expresses the fact that the 3-form $H$ is isometric, i.e.  the  two-form potential $B$ 
 transforms  by a gauge transformation under the isometry generated by $K_a$,
   $\mathcal L_K B=d(i_K B  -\XX)$, where $\mathcal L_K=i_K\, d+d\, i_K$ is the Lie derivative. It follows from \rf{222},\rf{22} that not only the three-form $H$ but also the one-form $\XX$ respect the isometry, i.e. 
      \be \la{225}  
      \mathcal L_K H=0    \ , \qquad \qquad      \mathcal L_K \XX=0  \ . \ee
  Furthermore, it follows from the ``Bianchi" part of the $\cal S$  equation \rf{414} 
  that the ``RR'' forms  in \rf{200} also respect the isometry
  \be\la{223}
	\mathcal L_K {\cal F}_{2n+1} = 0 \ , \ \ \ \ \ \ \ \   \ \ \ \ \   n=0,1,2 \ .   \ee
   Thus the whole bosonic  background  $(G,H,\mathcal F)$ is  $K$-isometric. This statement can,  in fact, be generalized to superspace as we will show in section  \ref{sec:lifting}.

  Assuming that  the  $B$-field  may be  chosen (by a gauge transformation) 
   to be isometric,  eq. \rf{22} 
    may be explicitly  solved as \cite{Arutyunov:2015mqj} (here $m,n$ are 10d coordinate indices)
 \be \la{2334}   \XX_m = \del_m \p   - B_{mn} K^n \ , \ \ \ \ \ \  {\rm i.e.} \ \ \ \ \ \ 
 X_m = \del_m \p    +  (G_{mn} - B_{mn}) K^n
 \ , \ee
 where  $\p$ is an arbitary scalar  that should  also satisfy the isometry condition 
 $K^m \del_m \p=0$   according to \rf{222}. 
 We conclude that   the  generalized system  of equations \rf{17}--\rf{20} 
  involves the standard fields of type II supergravity (including the  dilaton $\phi$) 
 plus an extra Killing vector $K_a$.\foot{Note that 
 if  the metric  admits several Killing vectors,  choosing  them as $K$   one by one  will lead to different  
 solutions of the generalized equations (e.g., different RR backgrounds).}

 Eqs. \rf{222} and \rf{22}  always  admit the following   special  solution 
\be K_a=0 \ , \ \ \qquad  \ \ \ \ \   X_a =\XX_a =\del_a \phi  \ . \la{23}\ee
 In that case the  generalized system \rf{17}--\rf{20} 
reduces to the  bosonic sector of the standard type IIB supergravity equations  
with $\p$  being  the dilaton.\foot{See, e.g.,  appendix A in \cite{Borsato:2016zcf} where the same RR bispinor notation for RR fields is used.}
In particular, eq.\rf{20} contains  both the  dynamical equations and the Bianchi identities  for the  RR  field strenghts $F_p = e^{-\p}{ \cal F}_p= d C_{p-1} +...$ in \rf{200}.

The above  generalized  type IIB   equations   have   of course 
a straightforward analog in type IIA case -- following from kappa-symmetry condition of  type IIA GS string.
Let us  also note    that there  is a natural generalization of the notion of a supersymmetric solution 
to the generalized type IIB supergravity equations, namely, the  one for which  the
component  fermionic fields as well as their supersymmetry variations vanish, i.e. $\chi_{\alpha i}|_{\theta=0}=\psi_{ab}^{\alpha i}|_{\theta=0}=0$ and $\epsilon^{\alpha i}\nabla_{\alpha i}\chi_{\beta j}|_{\theta=0}=\epsilon^{\alpha i}\nabla_{\alpha i}\psi_{ab}^{\beta j}|_{\theta=0}=0$. The latter two equations give the generalization of the dilatino and the (integrability\footnote{The Killing spinor equation itself takes the form (cf. \rf{410})
$
(\nabla_a+\frac18H_{abc}\,\gamma^{bc}\sigma^3+\frac18\mathcal S\gamma_a)\epsilon=0\,.
$
From the GS   sigma model perspective this condition follows from  the requirement of a  residual global 
supersymmetry of the action describing a  superstring moving in a non-trivial bosonic background. 
}
 of  the)  gravitino conditions respectively.  Using \rf{eq:nabla-chi-IIB} and \rf{eq:nabla-psi-IIB} they take the form
\begin{align}\te
&\te \big[
(\mathrm X_a+\sigma^3K_a)\gamma^a
+\frac{1}{12}H_{abc}\,\sigma^3\gamma^{abc}
+\frac{1}{8}\gamma_a\mathcal S\gamma^a
\big]\epsilon=\,0\,,
\\
\te
&\te \big[
R_{ab}{}^{cd}\gamma_{cd}
+\frac12H_{ace}H_{bd}{}^e\,\gamma^{cd}
-\nabla_{[a}H_{b]cd}\,\sigma^3\gamma^{cd}
-\nabla_{[a}\mathcal S\gamma_{b]}\qquad&
\nonumber\\
\te
&\te \qquad -\frac{1}{8}(\mathcal S\sigma^3\gamma_{[a}\gamma^{cd}-\gamma^{cd}\sigma^3\mathcal S\gamma_{[a})H_{b]cd}
-\frac{1}{8}\mathcal S\gamma_{[a}\mathcal S\gamma_{b]}
\big]\epsilon
=\,0\,.
\end{align}
These differ from the standard type IIB supersymmetry conditions only by the replacements $\nabla_a\phi\rightarrow\mathrm X_a+\sigma^3K_a$ and $e^{\phi}F\rightarrow\mathcal F$ inside the RR bispinor $\mathcal S$. It would be interesting to find solutions to these equations with $K_a\neq0$.
 
The   generalized  equations  \rf{17}--\rf{22} are precisely the ones identified in 
\cite{Arutyunov:2015mqj} as  being satisfied by the target space background 
of the so-called $\eta$-deformation   \cite{Klimcik:2008eq,Delduc:2013qra,Arutyunov:2015qva}
of the $AdS_5\times S^5$ superstring model.\foot{The relation to  the notation in  \cite{Arutyunov:2015mqj}  is  
 $\XX_a=Z_a$ and $K_a=I_a$.
As  in \cite{Arutyunov:2015mqj},  we find that while the NS-NS   subset of equations 
depends on $\XX_a$ and $K_a$ only through  their sum  $X_a$ in \rf{21}, the two   vectors enter separately 
in the RR equations  \rf{20}. While in the NS-NS   sector one does not need the orthogonality condition  $\XX_a K^a=0$, this condition  was, in fact,  imposed in \cite{Arutyunov:2015mqj}
once  the RR fields  were included  (see  eq. (5.37) there).}
The resulting picture is thus  in perfect  agreement 
with the  fact that the  $\eta$-model  is kappa-symmetric   \cite{Delduc:2013qra}
but  the corresponding  background does  not satisfy the type 
IIB equations \ci{Arutyunov:2015qva}.\footnote{This assumes that the action and kappa-symmetry transformations of the $\eta$-model are the same as those of the Green-Schwarz string. This   can 
 be shown to be the case  and  is also true for the $\lambda$-model of \cite{Hollowood:2014qma}
upon  integrating out the  superalgebra-valued  2d gauge field.}
Further  examples of solutions of the generalized  type  IIB equations  \rf{17}--\rf{22} 
should be   provided   by some other    $\eta$-models  \ci{Kyono:2016jqy,HS},  as was 
indeed shown in \ci{HS}  for the models  based on Jordanian R-matrices.



The solution \rf{23}  is   the only  possible one if the metric does not admit Killing vectors, i.e. 
kappa-symmetric GS sigma models  with  non-isometric metric   must correspond to 
standard type IIB solutions.  An example is provided   by the  $\lambda$-deformed  model 
which has kappa-symmetric   action \cite{Hollowood:2014qma} with the  corresponding  metric \ci{st} 
not admitting any Killing vectors:  as  was explicitly  demonstarted in \cite{Borsato:2016zcf} 
in the  $AdS_2\times S^2\times T^6$   case the  corresponding  $\lambda$-deformed 
background   solves the standard type IIB equations.

It  was argued in \cite{Arutyunov:2015mqj}  that  the above  generalized  type IIB 
equations  
 imply the scale-invariance conditions for the GS  sigma model.
 In particular, the 
  2nd-derivative scale-invariance conditions for the ``RR" fields
   follow immediately  upon  ``squaring"  of  the Dirac equation for $\mathcal S$ in \rf{20}.\foot{In general, the  equations  \rf{17}--\rf{20}  are thus somewhat stronger than the scale  invariance conditions, 
   but  still not sufficient to imply  the  Weyl invariance  unless $K_a=0$.}
Thus non-trivial  solutions  of the generalized equations with $K_a\not=0$ should represent UV finite 
but not Weyl-invariant GS sigma models so 
 their  string theory interpretation is  a priori unclear. 

As follows from the analysis in  \cite{Arutyunov:2015mqj},  starting with 
 a type IIA  supergravity solution 
with all  the  fields  being isometric  apart from a  linear term in the dilaton \ci{ht2},  
 and    performing  the standard 
   T-duality transformation on all the fields except the dilaton (i.e. on 
  the GS sigma model on a flat 2d background)\foot{In general,    T-duality of GS  sigma model on a flat 2d background  should preserve its kappa-symmetry  
   \cite{Kulik:2000nr}  and should  be expected 
   not take one out of 
  the class of solutions of the generalized  equations.}
   then the resulting background  should 
solve  precisely  the generalized equations \rf{17}--\rf{22}  with $K_a$ and $\XX_a$   determined by the original  dilaton and  the metric.\foot{In  particular, the resulting 
$K_a$ is then proportional to the derivative of the  dilaton along the non-isometric direction
 \cite{Arutyunov:2015mqj}.}
The converse should also  be true  \cite{Arutyunov:2015mqj} 
given a non-trivial  solution   of  the  generalized type IIB  equations \rf{17}--\rf{22} with a
non-null Killing vector $K_a$,  
  its $(G,B,\mathcal F)$ fields   should be related by
a T-duality transformation to   the  fields of  the corresponding type IIA supergravity solution with the
dilaton containing a linear isometry-breaking term. 
Thus each solution of the generalized type II system  \rf{17}--\rf{22}  can be
associated with a particular solution of the  standard type II supergravity equations. 
This  observation may 
help  understanding  if it is possible to  relate  a solution of the generalized type II  equations
  with  a consistent  string theory.


We shall   start in  section  2  with a derivation of the  type I   and type IIB  constraints \rf{1.1}  and \rf{1.6} 
on the superspace torsion and 3-form $H$   that follow from the  condition of kappa-symmetry for the GS   superstring in a non-trivial background. 
The solution of the Bianchi identities  supplemented  by  these basic constraints 
   leads, as  will be described 
in section 3, 
 to the generalized equations of motion \rf{1.3}--\rf{1.5}  and \rf{17}--\rf{22}.
In section 4 we shall present a superspace formulation of the equations on $K_a$ and $\XX_a$
and  invariance conditions and superspace Bianchi identities for the ``RR" form fields. 
Some concluding remarks  will be made in section 5.
Details of the solution of the superspace  Bianchi identities  will   be provided in an Appendix.

\section{Constraints from kappa-symmetry}\label{sec:kappa}

The classical GS superstring action in an arbitrary super-background 
is (in the Nambu-Goto form) \cite{Grisaru:1985fv}
\begin{equation}
S=\int d^2\xi\,\sqrt{-G}-\int_\Sigma\,B\,,\qquad \qquad G=\det{G_{IJ}}\,, \la{2.1}
\end{equation}
where $\xi^I$ ($I,J=0,1$)  are worldsheet coordinates,  $G_{IJ}$ is the induced metric
\begin{equation} \la{2.2}
G_{IJ}=E_I{}^aE_J{}^b\eta_{ab}\,,\qquad E_I{}^A=\partial_Iz^ME_M{}^A (z)\,,\qquad z^M=(x^m,\,\theta^\mu)\,,
\end{equation}
while $B$ is the pull-back of a superspace two-form. This action is required to be invariant under the following kappa-symmetry transformations of the coordinates $z^M$
\begin{equation} \la{2.3}\te 
\delta_\kappa z^ME_M{}^a=0\,,\qquad\delta_\kappa z^ME_M{}^{\alpha i}=\frac12(1+\Gamma)^{\alpha i}{}_{\beta j}\kappa^{\beta j}\,,\qquad\Gamma=\frac{1}{2\sqrt{-G}}\varepsilon^{IJ}E_I{}^aE_J{}^b\gamma_{ab}\sigma^3\,,
\end{equation}
where we have written the expressions appropriate to type IIB superspace. In the type IIA case  the Pauli matrix $\sigma^3$ is replaced by $\Gamma_{11}$ while in the type I case one is to  keep only the $i=1$ component. 
 The  operator   $\Gamma$ is traceless and  satisfies the projector condition 
 $\Gamma^2=1$ so 
  this symmetry removes half of the fermionic components.\foot{The origin of 
 kappa-symmetry is best understood via  embedding a worldvolume superspace in a target superspace. This so-called superembedding formalism is reviewed in \cite{Sorokin:1999jx}.}

The requirement that the string action be invariant under the above transformations imposes constraints on the background. We will now determine what these basic constraints are and in the next section we will work out all their consequences. Varying the action we find
\begin{equation} \la{2.4}
\delta_\kappa S=-\int d^2\xi\,\delta_\kappa z^ME_M{}^{\alpha i}
\left[
\sqrt{-G}\, G^{IJ}E_I{}^aE_J{}^CT_{C\alpha i}{}^b\eta_{ab}
\te  +\frac12\varepsilon^{IJ}E_I{}^CE_J{}^B H_{BC\alpha i}
\right]\,,
\end{equation}
where $T^A=dE^A+E^B\wedge\Omega_B{}^A$ is torsion   and 
$H=dB$. Note that the term involving the super-connection $\Omega_B{}^A$ 
does not contribute to \rf{2.4} 
 due to it being valued in $SO(1,9)$  (i.e. 
  $\Omega_{\alpha i}{}^b=0$ and $\Omega^{ab}=\Omega^{[ab]}$), it is nevertheless convenient to write the kappa-symmetry conditions  covariantly in terms of $T^A$ rather than $dE^A$. Since the (pulled-back) supervielbeins are assumed to be independent fields and since the projector $\Gamma$  only involves the bosonic supervielbeins, 
   eq. \rf{2.4}  implies  the following conditions 
\begin{align} \la{2.5}
&\qquad \qquad \qquad H_{\beta j\gamma k\alpha i}(1+\Gamma)^{\alpha i}{}_{\delta l}
=0\,,
\\ \  &\te \qquad
E_I{}^a
\left[
\sqrt{-G}\, G^{IJ}T_{\alpha i\beta j a}
-\varepsilon^{IJ}H_{a\alpha i\beta j}
\right](1+\Gamma)^{\alpha i}{}_{\gamma k}
=0\,,
\label{2.6}
\\ &\te \la{2.7}
E_I{}^aE_J{}^b\left[
\sqrt{-G}\,G^{IJ}T_{\alpha iab}
+\frac12\varepsilon^{IJ} H_{\alpha iab}
\right](1+\Gamma)^{\alpha i}{}_{\beta j}
=0\,.
\end{align}
The third condition turns out to be implied by the first  two, see eq. (\ref{eq:dim-half-IIB}) in the next section.

Since the two terms in   \rf{2.5} come with different powers of the induced metric, and since the components of $H$ cannot depend on the induced metric if $H$ is to have a target space interpretation, the this equation implies the vanishing of the dimension --$\ha$ component of the 3-form $H$
\begin{equation}\la{2.8}
H_{\alpha i\beta j\gamma k}=0\,.
\end{equation}
To solve the second condition \rf{2.6}  for the dimension 0  torsion and 3-form   components we parametrize these as
\begin{align}
T_{\alpha i\beta j}{}^a=
s^1_{ij}\gamma^b_{\alpha\beta}t_b{}^a
+s^2_{ij}\gamma^{bcdef}_{\alpha\beta}t_{bcdef}{}^a
+\varepsilon_{ij}\gamma^{bcd}_{\alpha\beta}t_{bcd}{}^a\,, \la{2.9}
\\
H_{a\alpha i\beta j}=
s^3_{ij}\gamma^b_{\alpha\beta}h_{ba}
+s^4_{ij}\gamma^{bcdef}_{\alpha\beta}h_{bcdefa}
+\varepsilon_{ij}\gamma^{bcd}_{\alpha\beta}h_{bcda}\,,\la{2.10}
\end{align}
where $s^{p}_{ij}$ are constant symmetric matrices and $t$  and $h$  are tensor superfields.
Tracing (\ref{2.6}) with $\gamma^{a}$ and multiplying with $G_{JK}$ we find the condition
\begin{align}
&  \te 
s^1_{ij}\sqrt{-G}E_K{}^bt_{ab}
-(s^3\sigma^3)_{ij}\frac{(\varepsilon G)^L{}_K\varepsilon^{IJ}}{\sqrt{-G}}E_L{}^cE_I{}^bE_{Ja}h_{bc}
-3\sigma^1_{ij}\frac{(\varepsilon G)^L{}_K\varepsilon^{IJ}}{\sqrt{-G}}E_L{}^dE_I{}^bE_J{}^ch_{abcd}
\nonumber\\
&{}
-s^3_{ij}\varepsilon^{IJ}G_{JK}E_I{}^bh_{ab}
+(s^1\sigma^3)_{ij}\varepsilon^{IJ}E_K{}^cE_I{}^bE_{Ja}t_{bc}
+3\sigma^1_{ij}\varepsilon^{IJ}E_K{}^dE_I{}^bE_J{}^ct_{abcd}
=0\,.
\label{2.6-1}
\end{align}
The first three terms and last three terms here  have to cancel independently since they come with different powers of the bosonic supervielbeins. The requirement that the last three terms cancel gives
\begin{equation}\la{2.12}
\varepsilon^{IJ}E_I{}^bE_J{}^cE_K{}^d
\left(
s^3_{ij}h_{ab}\eta_{cd}
+(s^1\sigma^3)_{ij}\eta_{ab}t_{cd}
-3\sigma^1_{ij}t_{abcd}
\right)
=0\,,
\end{equation}
implying that\foot{Note that the part anti-symmetric in $[bcd]$ vanishes trivially due to  the fact that the world-sheet indices $I,J,K$ only take two values.}
\begin{equation}\la{2.13}
s^3_{ij}h_{a[b}\eta_{c]d}
+(s^1\sigma^3)_{ij}(\eta_{a[b}t_{c]d}-\eta_{a[b}t_{cd]})
-3\sigma^1_{ij}(t_{abcd}-t_{a[bcd]})
=0\,.
\end{equation}
After  a little bit of algebra the solution is found to be
\begin{equation}\la{2.14}
s^3=s^1\sigma^3\,,\qquad t_{ab}=-h_{ab}=-i\eta_{ab}\,,\qquad t_{abcd}=t_{[abcd]}\,,
\end{equation}
where we have used the freedom to rescale the fermionic supervielbeins to normalize $t_{ab}$. The same freedom allows us to set $s^1_{ij}=\delta_{ij}$. The vanishing of the first three terms in (\ref{2.6-1}) then gives also
\begin{equation}
h_{abcd}=h_{[abcd]}\,.\la{2.15}
\end{equation}
Tracing (\ref{2.6}) with $\gamma^{abc}$ and $\gamma^{abcdef}$ and using the above conditions we find also that the components of  $t$ and $h$ fields  with more than two indices must vanish.

We conclude therefore that  the 
kappa-symmetry of the  type IIB GS string action  implies, in addition to \rf{2.8}, 
 the standard dimension 0 superspace  constraints
\begin{equation}\la{2.16}
T_{\alpha i\beta j}{}^a=-i\delta_{ij}\gamma^a_{\alpha\beta}\,,\qquad\qquad 
H_{a\alpha i\beta j}=-i\sigma^3_{ij}(\gamma_a)_{\alpha\beta}\,.
\end{equation}
The type IIA cases  can be analysed similarly. 
The constraints in the type I 
case \rf{1.1}  are  obtained by keeping only the $i=j=1$ components in
the  type IIB ones. 

The next step is to determine the consequences of these constraints by solving the superspace Bianchi identities for the torsion and  the 3-form $H$. This will lead us to the generalized supergravity equations described in the Introduction.


\section{Generalized equations 
 from   Bianchi identities and  constraints}\label{sec:solution}

Our aim will be to find the most general solution to the 10d  superspace Bianchi identities for the torsion and 3-form consistent with the dimension $-\ha$ \rf{2.8} 
    and  dimension 0  \rf{2.16}  constraints
following from kappa-symmetry of the GS string. 
We will consider the type I and type IIB cases in parallel
and present the summary of the results  while   details  will  be provided in the Appendix.

Let us first  recall the basic superspace conventions we will need.
The torsion satisfies the Bianchi identity\footnote{Our conventions are such  that $d$ acts from the right and the components of forms are defined as 

\noindent
$\omega^{(n)}=\frac{1}{n!}E^{A_n}\wedge\cdots\wedge  E^{A_1}\omega_{A_1\cdots A_n}$.}
\begin{equation}\nabla T^A=E^B\wedge R_B{}^A\,, \qquad \qquad 
T^A=\nabla E^A\equiv dE^A+E^B\wedge\Omega_B{}^A\,,\label{3.1}
\end{equation}
where  $R_B{}^A$ is the curvature superfield 2-form
\be \la{31}
R_B{}^A=d\Omega_B{}^A+\Omega_B{}^C\wedge\Omega_C{}^A
\ , \qquad \qquad \nabla R_B{}^A=0 \ .
\ee
 As follows from the fact that the 
 structure group is $SO(1,9)$,   the  non-zero  components  of the curvature 
 are  $R_a{}^b$ and
\begin{equation}\te 
\mbox{\bf Type I:}\quad R_\alpha{}^\beta=-\frac14R^{ab}(\gamma_{ab})^\beta{}_\alpha\,,\qquad\qquad 
\mbox{\bf Type IIB:}\quad R_{\alpha i}{}^{\beta j}=-\frac14R^{ab}\delta_{ij}(\gamma_{cd})^\beta{}_\alpha\,.
\label{eq:T0}
\end{equation}
In components,  the torsion  and curvature Bianchi identities in \rf{3.1} and \rf{31}  take  the form
\begin{align}
&\nabla_{[A}T_{BC]}{}^D+T_{[AB}{}^ET_{|E|C]}{}^D=R_{[ABC]}{}^D\,, \label{eq:torsion-bianchi}\\
&\nabla_{[A}R_{BC]D}{}^E+T_{[AB}{}^FR_{|F|C]D}{}^E=0\,.
\label{eq:curvature-bianchi}
\end{align}
A useful fact is that the curvature Bianchi identities are a consequence of the torsion Bianchi identities.\footnote{The proof goes as follows \cite{Dragon:1978nf}. Taking the covariant derivative of (\ref{3.1}) gives $E^B\wedge\nabla R_B{}^A=0$ and using the fact that the indices belong to the structure group
$SO(1,9)$  this implies $E^b\wedge\nabla R_b{}^a=0$ and $(\gamma_{ab}E)^{\alpha i}\wedge\nabla R^{ab}=0$. Analyzing the components of these equations it is not hard to see that they imply  the curvature Bianchi identity $\nabla R_B{}^A=0$.} This means that we only need to solve the torsion Bianchi identities.

We also  have to solve  the Bianchi identity for the  3-form $dH=0$, or, in components,
\begin{equation}\te 
\nabla_{[A}H_{BCD]}+\frac32T_{[AB}{}^EH_{|E|CD]}=0\,.
\label{eq:H-bianchi}
\end{equation}
There is some freedom in how one presents 
 the constraints.  We will write them in essentially the same form as the type II constraints of \cite{Wulff:2013kga}, which is particularly simple, rather than,  for example, in the form of the type I constraints used in \cite{Nilsson:1981bn}. The details of the solution to the Bianchi identities are given  in  Appendix.
 \foot{In Appendix  we    analyze also a  more general case   when in the type I  case 
 one imposes  only the torsion constraint in \rf{1.1}.}
  We shall discuss  the consequences of the  Bianchi identities   and constraints 
 in order of increasing dimension of the component superfields. 


\noindent
{\bf Dimension --$\ha$}:

As we have seen above,   kappa-symmetry of the string implies the vanishing of the dimension --$\ha$ component \rf{2.8} of the 3-form, i.e.
\begin{equation}\la{3.7}
\mbox{\bf Type I:}\qquad H_{\alpha\beta\gamma}=0\qquad\qquad\qquad
\mbox{\bf Type IIB:}\qquad H_{\alpha i\beta j\gamma k}=0\,.
\end{equation}

\noindent
{\bf Dimension 0}:

Kappa-symmetry of the string also requires the standard dimension 0  torsion and  3-form constraints \rf{2.16}
\begin{align}
&\la{3.8} \mbox{\bf Type I:}\qquad \quad  T_{\alpha\beta}{}^a=-i\gamma^a_{\alpha\beta}\,,\qquad\quad H_{a\alpha\beta}=-i(\gamma_a)_{\alpha\beta}\ , 
\\
&\la{3.9} \mbox{\bf Type IIB:}\qquad T_{\alpha i\beta j}{}^a=-i\delta_{ij}\gamma^a_{\alpha\beta}\,,\quad H_{a\alpha i\beta j}=-i\sigma^3_{ij}(\gamma_a)_{\alpha\beta}\,.
\end{align}
These are consistent with the dimension 0 Bianchi identity and the vanishing of the dimension --$\ha$ component of $H$.

\noindent
{\bf Dimension $\ha$}:

Let us  start with  the type I case. For the torsion we  shall  require that
\begin{equation}\la{3.10}
T_{\alpha[bc]}=0\ , 
\end{equation}
which just serves to fix the corresponding component of the spin connection, $\Omega_\alpha{}^{bc}$. By redefining the frame fields we can also arrange that\footnote{Taking
$E'=E+ i u E^b\gamma_cT_b{}^c+i v E^b\gamma_bT_c{}^c$ gives $T'_{(bc)}=T_{(bc)}-u\gamma_{(c}\gamma_dT_{b)}{}^d-v \eta_{bc}T_a{}^a$. This  implies $\gamma^bT'_{(bc)}=(1-6u)\gamma^bT_{(bc)}+(\ha u -v)\gamma_cT_b{}^b$ which vanishes for a suitable choice of  the constants $u,v$.
}
\begin{equation}
(\gamma^b)^{\alpha\beta}T_{\beta bc}=0\,.
\end{equation}
The torsion Bianchi identity we have to solve reads
\begin{equation}
T_{(\alpha\beta}{}^\delta\gamma^d_{\gamma)\delta}-\gamma^e_{(\alpha\beta}T_{\gamma)e}{}^d=0\,,
\end{equation}
where we used the form of the dimension 0 torsion component. 
With some work one can show that this, together with the Bianchi identity for the three-form, finally 
 implies
\begin{equation}\la{3.14}
\mbox{\bf Type I:}\qquad H_{\alpha bc}=0\,, \qquad 
 T_{\alpha b}{}^c=0\,,\qquad T_{\alpha\beta}{}^\gamma=2\delta^\gamma_{(\alpha}\chi_{\beta)}-\gamma_{\alpha\beta}^a(\gamma_a\chi)^\gamma\,,
\end{equation}
where  $\chi_\alpha$ is  some   MW spinor superfield.

For the type IIB case a similar analysis gives the following conditions
\begin{align}
\label{eq:dim-half-IIB} \mbox{\bf Type IIB:}\quad 
&T_{\alpha ib}{}^c=0\,,\qquad 
H_{\alpha i bc}=0\,,
\\
&\te  T_{\alpha i\beta j}{}^{\gamma k}=
\delta^{\gamma k}_{(\alpha i}\chi_{\beta j)}
+(\sigma^3\delta)^{\gamma k}_{(\alpha i}(\sigma^3\chi)_{\beta j)}
-\frac12\delta_{ij}\gamma_{\alpha\beta}^a(\gamma_a\chi)^{\gamma k}
-\frac12\sigma^3_{ij}\gamma_{\alpha\beta}^a(\gamma_a\sigma^3\chi)^{\gamma k}\,,
\nonumber
\end{align}
where  $\chi_{\alpha i}$ are  some  two MW  spinor superfields.

\noindent
{\bf Dimension 1}:

We shall impose the standard requirement 
\begin{equation}
T_{ab}{}^c=0\ , 
\end{equation}
which fixes the remaining components  $\Omega_c{}^{ab}$ of the spin connection. The type I torsion Bianchi identities we need to solve are then 
\be
-2iT_{c(\alpha}{}^\gamma\gamma^d_{\beta)\gamma}=R_{\alpha\beta c}{}^d
\ , \qquad \qquad 
\nabla_{(\alpha}T_{\beta\gamma)}{}^\delta
+T_{(\alpha\beta}{}^\epsilon T_{\gamma)\epsilon}{}^\delta
-i\gamma^a_{(\alpha\beta}T_{|a|\gamma)}{}^\delta
=R_{(\alpha\beta\gamma)}{}^\delta\,.
\ee
The Bianchi identity for the 3-form imposes the condition
\begin{equation}
T_{\alpha\beta}{}^eH_{cde}
+2T_{c(\alpha}{}^\gamma H_{\beta)\gamma d}
-2T_{d(\alpha}{}^\gamma H_{\beta)\gamma c}
=0\,.
\end{equation}
After some algebra one  obtains  the solution as
\begin{equation}\te
T_{a\alpha}{}^\delta=
\frac18(\gamma^{bc})_\alpha{}^\delta H_{abc}
\,,\qquad\qquad 
R_{\alpha\beta}{}^{cd}=\frac{i}{2}(\gamma_b)_{\alpha\beta}H^{bcd}
\,.
\end{equation}
In addition,  one finds that the derivative of the  spinor superfield $\chi$ in \rf{3.14} should be given by
\begin{equation}\mbox{\bf Type I:}\qquad \qquad\qquad 
\te
\nabla_\alpha\chi_\beta=\chi_\alpha\chi_\beta+\frac{i}{2}\gamma_{\alpha\beta}^aX_a-\frac{i}{24}\gamma^{abc}_{\alpha\beta}H_{abc}\,,
\label{eq:nablachi}
\end{equation}
where $X_a$ is some   vector superfield.\foot{While   as  superfields $\chi_\alpha$ and 
$X_a$ are of course  related, their first components   will be independent fields   entering the 
 generalized   equations. We will use same  notation for superfields and their lowest  components, with the  interpretation being  hopefully clear from  the context.}

In the type IIB case we find, by a similar analysis,
\begin{align}\te \mbox{\bf Type IIB:}\qquad 
&\te T_{a\alpha i}{}^{\delta j}=
\frac18(\gamma^{bc}\sigma^3)_{\alpha i}{}^{\delta j} H_{abc}
+\frac18(\gamma_a\mathcal S)_{\alpha i}{}^{\delta j}\,,\\ &
\te R_{\alpha i\beta j}{}^{cd}=\frac{i}{2}(\gamma_b\sigma^3)_{\alpha i\beta j}H^{bcd}-\frac{i}{4}(\gamma^{[c}\mathcal S\gamma^{d]})_{\alpha i\beta j}\,, 
\label{eq:T1-IIB}
\\
&\te\nabla_{\alpha i}\chi_{\beta j}=
\frac12\chi_{\alpha i}\chi_{\beta j}
+\frac12(\sigma^3\chi)_{\alpha i}(\sigma^3\chi)_{\beta j}
+\frac{i}{2}\gamma_{\alpha\beta}^a (\delta_{ij}\mathrm X_a   + \sigma^3_{ij} K_a)
\nonumber\\
&{}\te \qquad \qquad \qquad \qquad 
-\frac{i}{24}\sigma^3_{ij}\gamma^{abc}_{\alpha\beta}H_{abc}
-\frac{i}{16}(\gamma_a\mathcal S\gamma^a)_{\alpha i\beta j}\,,
\label{eq:nabla-chi-IIB}
\end{align}
where $\mathrm X_a$ and $K_a$ are  some vector superfields.
  $\mathcal S= (\mathcal S^{\alpha i, \beta j})$ is an anti-symmetric $32\times32$ matrix which is off-diagonal in $i,j$ and can therefore be represented as
\begin{equation}\la{324}\te
\mathcal S=-i\sigma^2\gamma^a\mathcal F'_a-\frac{1}{3!}\sigma^1\gamma^{abc}\mathcal F'_{abc}-\frac{1}{2\cdot 5!}i\sigma^2\gamma^{abcde}\mathcal F'_{abcde}\,,
\end{equation}
for  some $p$-form superfields $\mathcal F'_p$.\foot{The reason for the primes on ${\cal F}_p$ will become  clear in the next section
(the lowest   components  of 
$\mathcal F'_p $ and $\mathcal F_p $  will differ  only by  bilinear  fermionic terms).}

\noindent
{\bf Dimension $3\ov 2$}:	

The type I  torsion Bianchi identities to solve at dimension $3\ov 2$ are
\begin{align}
&-i\gamma^d_{\alpha\beta}T_{bc}{}^\beta=2R_{\alpha[bc]}{}^d
\ , \\
&\nabla_aT_{\beta\gamma}{}^\delta
-2\nabla_{(\beta}T_{|a|\gamma)}{}^\delta
+2T_{a(\beta}{}^\epsilon T_{\gamma)\epsilon}{}^\delta
-T_{\beta\gamma}{}^\epsilon T_{a\epsilon}{}^\delta
-i\gamma^e_{\beta\gamma}T_{ea}{}^\delta
=2R_{a(\beta\gamma)}{}^\delta\,. \la{326}
\end{align}
The first one is easily solved for the curvature as
\begin{equation}\te 
R_{\alpha bcd}=\frac{i}{2}(\gamma_b\psi_{cd})_\alpha-i(\gamma_{[c}\psi_{d]b})_\alpha\,,
\end{equation}
where $T_{ab}{}^\beta=\psi_{ab}^\beta$ is 
the gravitino field strength. Using this in  \rf{326}
one finds after a bit of algebra that the solution is
\begin{equation}\te 
 \mbox{\bf Type I:}\qquad
\nabla_\alpha H_{abc}
=
3i(\gamma_{[a}\psi_{bc]})_\alpha
\,,\qquad\qquad 
i(\gamma^b\psi_{ab})_\alpha
=
2\nabla_a\chi_\alpha
+\frac14(\gamma^{bc}\chi)_\alpha H_{abc}\,.
\label{eq:gammatracepsi}
\end{equation}
This solves the Bianchi identities but we must also remember the consistency conditions which follow from the equation for $\nabla_\alpha\chi_\beta$ in (\ref{eq:nablachi}). Taking another spinor derivative of this equation and symmetrizing we find an expression for the spinor derivative of $ X_a$
  \begin{align}\te 
\nabla_\alpha X_a
=&\te 
\frac12(\gamma^b\gamma_a\nabla_b\chi)_\alpha
+(\gamma_a\gamma^b\chi)_\alpha X_b
+\frac{1}{48}(\gamma_a\gamma^{bcd}\chi)_\alpha H_{bcd}
+\frac{1}{8}(\gamma^{bc}\chi)_\alpha H_{abc}\,.
\label{329}
\end{align}
A similar analysis in the  type IIB case gives the following superfield relations
\begin{align} 
 \mbox{\bf Type IIB:}\quad
&\te i(\gamma^b\psi_{ab})_{\alpha i}=2\nabla_a\chi_{\alpha i}+\frac14(\gamma^{bc}\sigma^3\chi)_{\alpha i}H_{abc}\,,\quad
\nabla_{\alpha i}H_{abc}=3i(\gamma_{[a}\sigma^3\psi_{bc]})_{\alpha i}\,,\quad
\label{eq:gravitino-eom}
\\
&\te R_{\alpha ibcd}=\frac{i}{2}(\gamma_b\psi_{cd})_{\alpha i}-i(\gamma_{[c}\psi_{d]b})_{\alpha i}\,,
  \\ \te
\nabla_{\alpha i}\mathcal S^{\beta1\gamma2}
\te =&
\mathcal S^{\beta1\gamma2}\chi_{\alpha i} \te
-2\delta_{\alpha i}^{[\beta1}(\mathcal S\chi)^{\gamma2]}
+2(\gamma^a\mathcal S)_{\alpha i}{}^{[\beta1}(\gamma_a\chi)^{\gamma2]}
+4i(\gamma^{ab})^{[\beta1}{}_{\alpha i}\psi_{ab}^{\gamma2]}\,,
\\
\nabla_{\alpha i}\mathrm X_a
=&\te
\nabla_a\chi_{\alpha i}
-\frac14(\gamma_a\gamma^b\nabla_b\chi)_{\alpha i}
+\frac12(\gamma_a\gamma^b
(\XX_b + \sigma^3K_b) \chi\big)_{\alpha i} 
\nonumber\\
&{}\te \qquad +\frac18(\gamma^{bc}\sigma^3\chi)_{\alpha i} H_{abc}
+\frac{1}{96}(\gamma_a\gamma^{bcd}\sigma^3\chi)_{\alpha i}H_{bcd}
+\frac{1}{16}(\gamma_a\mathcal S\chi)_{\alpha i} \ ,
\label{329-IIB}
\\
\nabla_{\alpha i}K_a
=&\te
-\frac14(\gamma_a\gamma^b\sigma^3\nabla_b\chi)_{\alpha i}
+\frac12\big(\gamma_a\gamma^b\sigma^3(\XX_b + \sigma^3K_b) \chi\big)_{\alpha i}
\nonumber\\
&{}\te\qquad 
+\frac{1}{96}(\gamma_a\gamma^{bcd}\chi)_{\alpha i}H_{bcd}
-\frac{1}{16}(\gamma_a\sigma^3\mathcal S\chi)_{\alpha i}\,.
\label{eq:nabla-K-IIB}
\end{align}

\noindent
{\bf Dimension 2}:

In the type I case the  torsion Bianchi identities read
\begin{equation} 
R_{[abc]}{}^d=0
\ , \qquad 
\nabla_\alpha T_{bc}{}^\beta
+2\nabla_{[b}T_{c]\alpha}{}^\beta
+2T_{[b|\alpha|}{}^\gamma T_{c]\gamma}{}^\beta
+T_{bc}{}^\gamma T_{\gamma\alpha}{}^\beta
=R_{bc\alpha}{}^\beta\,.
\end{equation}
They determine the spinor derivative of the gravitino field strength  superfield
\begin{equation}\la{336} 
\te 
\nabla_\alpha\psi_{ab}^\beta
=
\frac18(\gamma^{cd})^\beta{}_\alpha\big(2\nabla_{[a}H_{b]cd}+H_{eca}H^e{}_{bd}-2R_{abcd}\big)
-\delta^\beta_\alpha\psi_{ab}\chi
-\psi_{ab}^\beta\chi_\alpha
+(\gamma^c\psi_{ab})_\alpha(\gamma_c\chi)^\beta\,.
\end{equation}
We are finally ready 
to derive the  equations of motion for the bosonic superfields. 
Contracting \rf{336}  with $\gamma^a$ and using (\ref{eq:gammatracepsi}) gives the equations
\begin{align} &  \mbox{\bf Type I:}\qquad
\nabla_{[a}H_{bcd]}=0\,,\qquad \qquad R_{a[bcd]}=0\,,
\label{eq:H-R-bianchi} \\ &\te 
\nabla^cH_{abc}-4\nabla_{[a}X_{b]}-2X^cH_{abc}-4\psi_{ab}\chi=0\,,\qquad
R_{ab}+2\nabla_{(a}X_{b)}-\frac14H_{acd}H_b{}^{cd}=0\,,
\label{eq:einstein-eq-I}
\end{align}
where $R_{ab}=R_{ac}{}^c{}_b$. 
 Evaluating $\nabla_{(\alpha}\nabla_{\beta)}X_a$  and 
using (\ref{329}) we find  also 
\begin{equation}\te
\nabla^aX_a-2X^aX_a+\frac{1}{12}H^{abc}H_{abc}+2i\chi\gamma^a\nabla_a\chi-\frac{i}{12}\chi\gamma^{abc}\chi\,H_{abc}=0\,.
\label{eq:div-X-I}
\end{equation}
The lowest components of these superfield equations 
give us   the generalized type I equations \rf{1.3}--\rf{1.5}  discussed
 in the Introduction (where fermionic components  were set to zero).

In the type IIB  case  one finds the fermionic equation 
(\ref{eq:nabla-psi-IIB}) together with the following equations  for the bosonic superfields\foot{Here the covariant derivatives (e.g. in \rf{eq:dh-etc})
contain fermionic terms so,  e.g.,  $K_m=0, \ \XX_m = \del_m \phi$ is always a solution  even for non-zero fermionic   fields.}
\begin{align}
&
  \mbox{\bf Type IIB:}\qquad
R_{a[bcd]}=0\,,\qquad\nabla_{[a}H_{bcd]}=0\,,\la{eq:dH}\\
&
\qquad
2\nabla_{[a}\mathrm X_{b]}+K^cH_{abc}+\psi_{ab}\chi=0\,,\qquad\nabla_{(a}K_{b)}=0\,,
\label{eq:dh-etc}
\\
&\qquad \te K^a\mathrm X_a-\frac{i}{4}\chi\gamma^a\sigma^3\nabla_a\chi+\frac{i}{96}\chi\gamma^{abc}\chi\,H_{abc}=0\,, 
\label{eq:div-K}
\\
&\te \qquad R_{ab}+2\nabla_{(a}\mathrm X_{b)}-\frac14H_{ade}H_b{}^{de}+\frac{1}{128}\mathrm{Tr}(\mathcal S\gamma_a\mathcal S\gamma_b)=0\,,
\label{eq:einstein-eq-IIB}
\\
&\qquad   \te \nabla^cH_{abc}-2{\mathrm X}^cH_{abc}-4\nabla_{[a}K_{b]} -\frac{1}{64}\mathrm{Tr}(\mathcal S\gamma_a\mathcal S\gamma_b\sigma^3)-2\psi_{ab}\sigma^3\chi=0\,,
\label{eq:div-H}
\\
&\qquad \te \nabla^a\mathrm X_a-2{\mathrm X}^a\mathrm X_a-2K^aK_a+\frac{1}{12}H^{abc}H_{abc}\nonumber\\
& \qquad\qquad  \te -\frac{1}{256}\mathrm{Tr}(\mathcal S\gamma^a\mathcal S\gamma_a)+i\chi\gamma^a\nabla_a\chi-\frac{i}{24}\chi\gamma^{abc}\sigma^3\chi\,H_{abc}=0\,,
\label{eq:div-X-IIB}
\\
&\qquad \te
(\gamma^a\nabla_a\mathcal S)_{\alpha i}{}^{\beta j}
-\big(\gamma^a (\XX_a + \sigma^3 K_a) \mathcal S\big)_{\alpha i}{}^{\beta j}
+\big[\frac18(\gamma^a\sigma^3\mathcal S\gamma^{bc})_{\alpha i}{}^{\beta j}
+\frac{1}{24}(\gamma^{abc}\sigma^3\mathcal S)_{\alpha i}{}^{\beta j}\big]H_{abc}
\nonumber\\
&{}\qquad 
+i\chi_{\alpha i}(\mathcal S\chi)^{\beta j}
-i(\sigma^3\chi)_{\alpha i}(\sigma^3\mathcal S\chi)^{\beta j}
+2(\gamma^{cd}\chi)_{\alpha i}\psi_{cd}^{\beta j}
-2(\gamma^{cd}\sigma^3\chi)_{\alpha i}(\sigma^3\psi_{cd})^{\beta j}
=0\,.
\label{eq:RR-eom}
\end{align}
These are the generalized type IIB equations implied by the kappa-symmetry of the GS  string (generalizing \rf{17}--\rf{20}  where fermions were set to zero). 
One can show  that  they 
reduce to the standard type IIB supergravity equations in the special case of $K_a=0$.

\section{Lifting the Killing vector   and  IIB  form fields  to superspace}\label{sec:lifting}

The  generalized  type IIB equations in the previous section can be formulated in a geometrical way in superspace by lifting the Killing vector field $K_a$ and the
form fields  $\mathcal F_p$ to  superspace vector field and  superspace forms.
 We begin with the one-form   with 10d  coordinate components $\XX_m$ 
 and lift it to a one-form $\XX=dz^M\XX_M$  in superspace. We must then constrain the extra spinor
 component  not to introduce extra degrees of freedom.  This is  done  by  
  imposing  the constraint
\begin{equation}\la{4.1} 
\mathrm X_{\alpha i}=\chi_{\alpha i}\,.
\end{equation}
The equation for $\nabla_{[a}\mathrm X_{b]}$  in (\ref{eq:dh-etc})
 as well as the equation (\ref{329-IIB}) 
 for $\nabla_{\alpha i}\mathrm X_a$  and the equation for $\nabla_{(\alpha i}\chi_{\beta j)}$ in  (\ref{eq:nabla-chi-IIB}) are then all summarized by  the ``superspace Bianchi identity"
\begin{equation}
d\mathrm X+i_KH=0\qquad\Leftrightarrow\qquad\mathcal L_KB=d(i_KB-\mathrm X)\,,
\label{eq:X-bianchi}
\end{equation}
or,  in components, 
\begin{equation}\la{4.3}
2\nabla_{[A}\mathrm X_{B]}+T_{AB}{}^C\mathrm X_C=-K^CH_{ABC}\,.
\end{equation}
This equation says  that $B$ transforms by a gauge transformation under the superisometries generated by $K^A=(K^a,\Xi^{\alpha i})$, where the Killing spinor superfield $\Xi^{\alpha i}$ is set to be 
\begin{equation}\te \la{4.4}
\Xi
=
\frac{i}{4}
\big(\gamma^a\nabla_a
-2\gamma^a\,\mathrm X_a
-2\gamma^a\sigma^3\,K_a
-\frac{1}{24}\gamma^{abc}\sigma^3\,H_{abc}
-\frac{1}{4}\mathcal S
\big)\sigma^3\chi\,.
\end{equation}
This definition together with (\ref{eq:div-K}) implies that
\begin{equation}
i_K\mathrm X=0\,.
\label{eq:KX}
\end{equation}
Using (\ref{eq:X-bianchi}) we then conclude that  (super)isometries generated by $K$ leave $\mathrm X$ invariant  (the  rotation  matrix $L_A{}^B$ is defined below)
\begin{equation}
\mathcal L_K\mathrm X=0\,,
\ \ \ \qquad  {\rm i.e.} \ \ \ \ \qquad
K^C\nabla_C\mathrm X_A+L_A{}^B\mathrm X_B+i_K\Omega_A{}^B\mathrm X_B=0\,.
\end{equation}
Indeed,  the superspace  vector field $K^A$ satisfies the superspace Killing equation (see,  for example,  \cite{Wulff:2015mwa})
\begin{equation}\la{48} 
E^BL_B{}^A=\mathcal L_KE^A=\nabla K^A+i_KT^A-E^Bi_K\Omega_B{}^A\,.
\end{equation}
This equation expresses  the fact that
under the superisometry  generated by the vector superfield $K^A$
 the frame $E^A$ transforms by a local Lorentz transformation with the parameter $L_B{}^A=(L_b{}^a,\frac14L_{ab}(\gamma^{ab})_{\beta j}{}^{\alpha i})$ . 
The  component form of  \rf{48} is 
\begin{equation}\la{49}
\nabla_BK^A+K^CT_{CB}{}^A=L_B{}^A+i_K\Omega_B{}^A\,.
\end{equation}
Taking the parameter of the local Lorentz transformation to be
\begin{equation}
L_{ab}
=
\nabla_{[a}K_{b]}
-i_K\Omega_{ab}
\,,
\end{equation}
one  gets, from the $(ab)$ component of \rf{49},  
 the standard Killing vector equation $\nabla_{(a}K_{b)}=0$. The $(a\beta j)$-component gives the equation  (\ref{eq:nabla-K-IIB}) for $\nabla_{\alpha i}K_a$. The $(\alpha i\beta j)$ component implies the equation 
(\ref{eq:RR-eom})   for $\mathcal S$ (except for the $\gamma^{abcd}$ part), 
and also  the equation of motion  (\ref{eq:div-H}) for $B$, 
 the equation (\ref{eq:div-X-IIB})   for the divergence of $\mathrm X_a$, 
 as well as the constraint (\ref{eq:div-K}) 
 on $K^a\mathrm X_a$. To show  this requires  using the equation for the spinor derivative of the bosonic fields and the gravitino equation of motion. Finally,  the $(\alpha ib)$ component of \rf{49} is the superspace Killing spinor equation
\begin{equation}\te \la{410}
\nabla_b\Xi^{\alpha i}
+\frac18(\gamma^{cd}\sigma^3\Xi)^{\alpha i}\,H_{bcd}
+\frac18(\mathcal S\gamma_b\Xi)^{\alpha i}
-K^c\psi_{bc}{}^{\alpha i}
=0\,,
\end{equation}
and its  lowest component is the usual Killing spinor equation.\foot{This equation \rf{410} is not independent and arises by taking a spinor derivative of the $(\alpha i\beta j)$ component of \rf{49},  symmetrizing and using 
the  other equations  given above.}

Finally,  we can also lift  to superspace the form fields appearing in the bispinor 
$\mathcal S$ in \rf{324} 
setting there  
\begin{equation}
\mathcal F'_{a_1\cdots a_n}=\mathcal F_{a_1\cdots a_n}+i\chi^1\gamma_{a_1\cdots a_n}\chi^2\,.
\end{equation}
This works almost identically the same as for the standard type IIB  supergravity 
 theory 
where $\mathcal F_p$ are the RR field strengths multiplied by $e^\phi$ \cite{Wulff:2013kga}. Imposing the following constraints on their dimension 0 and dimension 
$\ha$ components
\begin{align}
&\qquad \mathcal F_{\alpha i\beta jc}=i\sigma^1_{ij}(\gamma_c)_{\alpha\beta}\,,\qquad\qquad 
\mathcal F_{\alpha i\beta jcde}=-\sigma^2_{ij}(\gamma_{cde})_{\alpha\beta}
\ , \\
&\mathcal F_{\alpha i}=-i(\sigma^2\chi)_{\alpha i}\,,\qquad
\mathcal F_{\alpha ibc}=-(\sigma^1\gamma_{bc}\chi)_{\alpha i}\,,\qquad
\mathcal F_{\alpha ibcde}=-i(\sigma^2\gamma_{bcde}\chi)_{\alpha i}\,,
\end{align}
one can show that they satisfy the following ``generalized Bianchi identities"
(same as  in  \ci{Arutyunov:2015mqj} for 10d  components)\foot{The $n=-1$ case corresponds to the condition $i_K \, \mathcal F_1=0$    \ci{Arutyunov:2015mqj}.}
\begin{equation}
d\mathcal F_{2n+1}+\mathrm X\wedge\mathcal F_{2n+1}-H\wedge\mathcal F_{2n-1}-i_K\mathcal F_{2n+3}=0\ , \qquad n=-1,0,1,2\,,\la{414}
\end{equation}
or,  in components, 
\begin{align}\te
\nabla_{[A_1}\mathcal F_{A_2\cdots A_{2n+2}]}
+\frac{2n+1}{2}T_{[A_1A_2}{}^B\mathcal F_{|B|A_3\cdots A_{2n+2}]}
-\mathrm X_{[A_1}\mathcal F_{A_2\cdots A_{2n+2}]}\nonumber
&\\  \te
+\frac{(2n+1)2n}{3!}H_{[A_1A_2A_3}\mathcal F_{A_4\cdots A_{2n+2}]}
-\frac{1}{2n+2}K^B\mathcal F_{BA_1\cdots A_{2n+2}}
\,&=0\,.
\end{align}
It is easy to check, using (\ref{eq:X-bianchi}) and (\ref{eq:KX}), that as a consequence of these generalized Bianchi identities the forms $\mathcal F_p$ are also invariant under the (super)isometries generated by $K$, i.e.
\begin{equation}
\mathcal L_K\mathcal F_{2n+1}=0\,, \ \ \ \ \ \ \ \ \ \ \ \ \ \ \  n=0,1,2\,.  \la{416}
\end{equation}

\section{Concluding remarks}\label{sec:conclusion}

In this paper we have   found  the equations imposed on the target space (super) geometry by the requirement that the classical Green-Schwarz superstring should 
 be kappa-symmetric. The bosonic part of these equations are exactly the same as suggested earlier in \ci{Arutyunov:2015mqj}.
  The   resulting  generalization of the standard 10d supergravity 
  equations  is  automatically supersymmetric as it was 
   obtained  from a superspace construction.  
    There is  also  a straightforward generalization of the notion of a supersymmetric solution of the generalized equations. 
    
    We have performed the detailed analysis  for the type I and type IIB cases but the corresponding generalized type IIA equations can be written down almost immediately using the results of  \cite{Wulff:2013kga}. 
   One open question  (raised already in \ci{Arutyunov:2015mqj}) is whether  these equations \rf{17}--\rf{20}   can be derived from an action and  should thus  satisfy certain   integrability conditions. 
  Another  is  about possible  uplift of the generalized  type IIA equations  to 11 dimensions 
and a  relation to a (partially off-shell?)  generalization of 11d supergravity.
    
  Non-trivial solutions of the type II  generalized equations 
  describe    backgrounds  symmetric   with respect to the vector $K_a$.  Applying T-duality 
  one then gets a type II  supergravity  solution with  a dilaton  containing a linear non-isometric term
   \ci{ht2,Arutyunov:2015mqj}. 
   It would  be interesting  to extend the discussion in \ci{Arutyunov:2015mqj}
   to determine   how  more  general T-dualities   act on these equations. 
   Applying T-duality to the GS  sigma model  \cite{Kulik:2000nr}
   should  transform  the background fields in a way consistent with kappa-symmetry  
 and should thus map  one solution of the generalized equations  to another.

 To investigate the properties  of the corresponding  sigma models 
 one  may  consider  the  component  expansion  of the type II GS 
  superstring action in these more general backgrounds.  
  This expansion  takes the same form  as in the standard type II supergravity backgrounds 
  \cite{Wulff:2013kga}
  provided one replaces the dilaton-modified RR field strengths $ e^\phi F_p$ by $\mathcal F_p$ and the dilaton gradient term $\frac{i}{2}\delta_{ij}\gamma^a\partial_a\phi$ in the  quartic fermion terms (appearing in the matrix $T$  in  \cite{Wulff:2013kga})
   by $\frac{i}{2}\gamma^a(\delta_{ij}\mathrm X_a+\sigma^3_{ij}K_a)$.
\vspace{.5cm}

\noindent
{\bf Note added:} After this paper appeared in arXiv we were informed of an earlier work on the pure spinor superstring that also observed that classical BRST invariance, the analog of kappa symmetry in that formulation, is not enough to restrict  the background to be a supergravity solution \cite{Mikhailov:2012id}. The relation with the condition $\chi_{\alpha i}=\nabla_{\alpha i}\phi$ and the fact that the generalized backgrounds (referred to there as ``non-physical'') are connected with global symmetries were commented on in section  7.3 of \cite{Mikhailov:2014qka}.

\section*{{Acknowledgments}}
We  acknowledge  R. Borsato, B. Hoare   and C. Hull   for useful discussions.
We  are grateful to R. Borsato, B. Hoare and R. Roiban   for   helpful comments on the draft. 
AAT  would like to thank R. Roiban for  important 
 discussions on  the relation between $\kappa$-symmetry   and supergravity constraints.
LW is grateful to P. Howe for important  clarifying discussions. 
We also thank A. Mikhailov for pointing out references \cite{Mikhailov:2012id,Mikhailov:2014qka} to us.
This work was supported by the ERC Advanced grant No.290456. 
The work of AAT   was  also supported by the 
STFC Consolidated grant  ST/L00044X/1 and 
 by the Russian Science Foundation grant 14-42-00047.


\appendix

\section*{Appendix}
\section{Details of solution of   superspace Bianchi identities\\  and constraints}\label{app:bianchi}

Here we will provide   details of the solution of the Bianchi identities for
the  torsion and the 3-form $H$ presented in section \ref{sec:solution}. 
The relevant Bianchi identities are (\ref{eq:torsion-bianchi}) and (\ref{eq:H-bianchi}).

We shall start from the constraints imposed by the kappa-symmetry on the dimension --$\ha$ and dimension 0 components   as found in  section  \ref{sec:kappa} \foot{To recall,  in this paper
$a,b=0,1,...,9; \ \ \  \alpha,\beta=1,2,...,16; \ \ \ i,j=1,2$.}
\begin{align}
\mbox{\bf Type I:}&\qquad H_{\alpha\beta\gamma}=0\,,\quad \qquad T_{\alpha\beta}{}^a=-i\gamma^a_{\alpha\beta}\,,\qquad \ \  H_{a\alpha\beta}=-i(\gamma_a)_{\alpha\beta}\ , \la{a1} \\
\mbox{\bf Type IIB:}&\qquad H_{\alpha i\beta j\gamma k}=0\,,\qquad T_{\alpha i\beta j}{}^a=-i\delta_{ij}\gamma^a_{\alpha\beta}\,,\quad H_{a\alpha i\beta j}=-i\sigma^3_{ij}(\gamma_a)_{\alpha\beta}\,.\la{a2}
\end{align}
We will proceed  by dimension of $T$ and $H$   components 
and  at each dimension  will  first work out the solution of the type I Bianchi identities and then present the type IIB solution. The type IIA solution should take an almost identical form to  type IIB one 
as is clear from the discussion in  \cite{Wulff:2013kga}.

In the type I case we will 
  be more general: 
  we   will first   impose only the dimension 0  constraint on the torsion
and   then comment  on  additional conditions following from  including the 3-form constraint
at the end of each subsection.
 In that case 
one obtains a more general solution which contains two 3-form fields which we call $g_{abc}$ and $h_{abc}$, see e.g. \cite{Howe:1986ed}. This more general version of type I supergravity is, of course, 
 not directly  relevant for   string theory  as   kappa-symmetry requires the presence of the 
 3-form $H$ satisfying  the above constraints.

\subsection*{Dimension $\ha$}

Starting with the {\bf type I} case, at  dimension $\ha$ we need to solve the torsion Bianchi identity
\begin{equation}
T_{(\alpha\beta}{}^\delta\gamma_{\gamma)\delta}^d-\gamma^e_{(\alpha\beta}T_{\gamma)e}{}^d=0\,,
\label{eq:dim-1/2-I}
\end{equation}
where $T_{\alpha[bc]}=0$ and $(\gamma^b)^{\alpha\beta}T_{\alpha b}{}^c=0$ (see section \ref{sec:solution})  and we used the dimension zero constraint 
on the torsion in \rf{a1}. Contracting with $\gamma_b^{\beta\gamma}$ this gives the equation
\begin{equation}
2(\gamma^d\gamma_b)_\delta{}^\beta T_{\alpha\beta}{}^\delta
+\gamma_b^{\beta\gamma}T_{\beta\gamma}{}^\delta\gamma_{\alpha\delta}^d
-20T_{\alpha b}{}^d=0\,.
\end{equation}
Expanding in a basis of gamma matrices
\begin{equation}
T_{\alpha\beta}{}^\gamma=\gamma_{\alpha\beta}^a\psi_a^\gamma+\gamma^{abcde}_{\alpha\beta}\psi_{abcde}^\gamma\,,
\end{equation}
this equation implies (using the symmetry and gamma-tracelessness of $T_{\alpha bc}$)
\begin{align}\te
&\te \gamma^{abcde}\psi_{abcde}=-\frac95\gamma^a\psi_a\,,\qquad
T_{\alpha ab}=
\frac45(\gamma_{(a}\psi_{b)})_\alpha
-\frac{2}{25}\eta_{ab}(\gamma^c\psi_c)_\alpha
\\
&\gamma^a\gamma_{fg}\psi_a
+\gamma^{abcde}\gamma_{fg}\psi_{abcde}
-8\gamma_{[f}\psi_{g]}
=0\,.
\end{align}
Multiplying the second equation with $\gamma^a$ and using the gamma-tracelessness of $T_{\alpha ab}$ we find
\begin{equation}\te 
\psi_a=-\frac{7}{8}\gamma_a\chi\,,
\end{equation}
for some spinor superfield $\chi$ whose normalization we have chosen for later convenience. Using this in the above equations we find
\begin{equation}\te
T_{\alpha a}{}^b=0\,,\qquad
\gamma^{abcde}\psi_{abcde}=\frac{63}{4}\chi\,,\qquad
\gamma^{abc}\psi_{abcfg}=\frac{7}{32}\gamma_{fg}\chi-\frac14\gamma_{[f}\gamma^{abcd}\psi_{g]abcd}\,.
\end{equation}
Contracting the dimension $\ha$ Bianchi identity (\ref{eq:dim-1/2-I}) with $\gamma_{ghabc}^{\beta\gamma}$ we get 
\begin{equation}\te 
16\cdot5!\gamma^f\psi_{ghabc}
+\gamma^{pqrde}\gamma_{ghabc}\gamma^f\psi_{pqrde}
-\frac{7}{4}\gamma^f\gamma_{ghabc}\chi
=0\,.
\end{equation}
Multiplying  this equation with $\gamma_f$ gives
\begin{equation}
\psi_{ghabc}\te 
=
\frac{7}{64\cdot5!}\gamma_{ghabc}\chi
-\frac{1}{16\cdot5!}\gamma^{pqrd}\gamma_{ghabc}\gamma^e\psi_{pqrde}\,.
\end{equation}
This equation determines $\psi_{abcde}$ recursively and after some algebra 
one finds
\begin{equation}\te 
\psi_{abcde}=\frac{1}{16\cdot5!}\gamma_{abcde}\chi\,.
\end{equation}
This completes the solution of the dimension $\ha$ torsion Bianchi identity. The non-vanishing torsion at dimension $\ha$ is thus 
\begin{equation}\te 
T_{\alpha\beta}{}^\gamma=
-\frac{7}{8}\gamma_{\alpha\beta}^a(\gamma_a\chi)^\gamma
+\frac{1}{16\cdot5!}\gamma^{abcde}_{\alpha\beta}(\gamma_{abcde}\chi)^\gamma
=
2\delta^\gamma_{(\alpha}\chi_{\beta)}
-\gamma_{\alpha\beta}^a(\gamma_a\chi)^\gamma\,.
\end{equation}
Imposing also the dimension $\ha$ Bianchi identity for the 3-form\foot{As usual,  $|E|$ means that index $E$ is not    symmetrized.}
\begin{equation}
3\nabla_{(\alpha}H_{\beta\gamma)d}
-\nabla_dH_{\alpha\beta\gamma}
+3T_{(\alpha\beta}{}^EH_{|E|\gamma)d}
-3T_{d(\alpha}{}^EH_{|E|\beta\gamma)}
=0
\end{equation}
and using the dimension 0 and dimension --$\ha$ constraints in \rf{a1} we get
\begin{equation}
\gamma^a_{(\alpha\beta}H_{\gamma)ab}=0\,,
\end{equation}
which implies the vanishing of the dimension $\ha$  component of $H$
\begin{equation}
H_{\alpha bc}=0\,.
\end{equation}


In the {\bf type IIB} case  the torsion  Bianchi identity is
\begin{equation}
T_{(\alpha i\beta j}{}^{\delta l}T_{\gamma k)\delta l}{}^d-T_{(\alpha i\beta j}{}^eT_{\gamma k)e}{}^d=0\,.
\end{equation}
When $i=j=k$ the analysis is the same as above and we get
\begin{align}
&\qquad \qquad \qquad T_{\alpha ib}{}^c=0
\ , \\
&T_{\alpha1\beta1}{}^{\gamma1}=
2\delta^\gamma_{(\alpha}\chi^1_{\beta)}
-\gamma_{\alpha\beta}^a(\gamma_a\chi^1)^\gamma
\,,\qquad
T_{\alpha2\beta2}{}^{\gamma2}=
2\delta^\gamma_{(\alpha}\chi^2_{\beta)}
-\gamma_{\alpha\beta}^a(\gamma_a\chi^2)^\gamma\,.
\end{align}
The remaining components of the Bianchi identity give 
\begin{equation}\la{a20}
T_{\alpha 1\beta 1}{}^{\delta 2}\gamma^d_{\gamma\delta}
+2T_{\gamma 2(\alpha 1}{}^{\delta 1}\gamma^d_{\beta)\delta}
=0\,,
\end{equation}
and the same equation with the indices $1$ and $2$ interchanged. 

From the dimension $\ha$ Bianchi identity for the 3-form  we get\footnote{If we do not 
 impose also the 3-form Bianchi identity there exists a much more general solution
\begin{equation}\nonumber
T_{\alpha i\beta j}{}^{\gamma k}=
2\delta_{(\alpha}^\gamma\Lambda^{ijk}_{\beta)}
-\gamma^a_{\alpha\beta}(\gamma_a\Lambda^{ijk})^\gamma
+2i(\sigma^2\delta)_{(\alpha i}{}^{\gamma k}\psi_{\beta j)}\,,
\end{equation}
where $\Lambda^{ijk}$ is a spinor superfield completely symmetric in the $SO(2)$ indices $ijk$ and $\psi$ is another spinor superfield.}
\begin{equation}
T_{(\alpha i\beta j}{}^{\delta l} H_{\gamma k)\delta l d} \la{a21}
=0\,.
\end{equation}
When $i=j=k$ this equation implies  the vanishing of the dimension $\ha$ component of the 3-form as before
\begin{equation}
H_{\alpha ibc}=0\,.
\end{equation}
The other  components of  \rf{a21} give 
\begin{equation}
T_{\alpha 1\beta 1}{}^{\delta 2}(\gamma_d)_{\gamma\delta}
-2T_{\gamma 2(\alpha 1}{}^{\delta 1}(\gamma_d)_{\beta)\delta}
=0\,,
\end{equation}
and the same with indices $1$ and $2$ interchanged. Together with eq. \rf{a20}
this  leads to the vanishing of the remaining components of the torsion.

\subsection*{Dimension 1}

The {\bf type I}  torsion Bianchi identities at dimension 1 read
\begin{align}
&-2iT_{c(\alpha}{}^\gamma \gamma_{\beta)\gamma}^d=R_{\alpha\beta c}{}^d
\label{eq:dim-1-I1}
\ , \\
&\nabla_{(\alpha}T_{\beta\gamma)}{}^\delta
+T_{(\alpha\beta}{}^\epsilon T_{\gamma)\epsilon}{}^\delta
-i\gamma^a_{(\alpha\beta}T_{|a|\gamma)}{}^\delta
=R_{(\alpha\beta\gamma)}{}^\delta\,,
\label{eq:dim-1-I2}
\end{align}
where we used the lower dimension constraints and the fact that $T_{ab}{}^c=0$. The first equation defines the curvature in terms of the torsion and using this in the second equation we find
\begin{equation}\te
\nabla_{(\alpha}T_{\beta\gamma)}{}^\delta
+T_{(\alpha\beta}{}^\epsilon T_{\gamma)\epsilon}{}^\delta
-i\gamma^a_{(\alpha\beta}T_{|a|\gamma)}{}^\delta
-\frac{i}{2}T_{a(\alpha}{}^\epsilon(\gamma_b)_{\beta|\epsilon|}(\gamma^{ab})^\delta{}_{\gamma)}
=0\,.
\end{equation}
Multiplying by $\gamma^c_{\eta\delta}$ and symmetrizing in $(\alpha\beta\gamma\eta)$ we get, using the dimension $\ha$ Bianchi identity,
\begin{equation}
T_{a(\alpha}{}^\delta\gamma^a_{\beta\gamma}\gamma^c_{\eta)\delta}=0\,.
\end{equation}
Let us now expand  $T_{a\alpha}{}^\delta$   in a basis of gamma matrices
\begin{equation}
T_{a\alpha}{}^\delta=
\delta_\alpha^\delta f_a
+(\gamma_{cd})_\alpha{}^\delta f_a{}^{cd}
+(\gamma_{cdef})_\alpha{}^\delta f_a{}^{cdef}\,.
\end{equation}
The first Bianchi identity \rf{eq:dim-1-I1}  implies, using the anti-symmetry of its r.h.s. 
 in $cd$ that
\begin{equation}\te
f_{(ab)c}=\frac12\eta_{c(a}f_{b)}\,,\qquad \qquad(\gamma_{cdef(a})_{\alpha\beta}f_{b)}{}^{cdef}=0\,.
\end{equation}
These conditions further imply
\begin{equation}\te
f_b{}^{ab}=\frac{11}{2}f^a\,,\qquad f_a{}^{cdef}=\frac{1}{48}\delta_a^{[c}g^{def]}\,,
\end{equation}
for some 3-form $g_{abc}$. Then 
\begin{equation}
T_{a(\alpha}{}^\delta\gamma^a_{\beta\gamma}\gamma^c_{\eta)\delta}=0
\qquad\Rightarrow\qquad
\gamma^b_{(\alpha\beta}\gamma^a_{\gamma\delta)}f_a=0\qquad\Rightarrow\qquad f_a=0\,.
\end{equation}
We therefore get
\begin{equation}\te
T_{a\alpha}{}^\delta=
\frac18(\gamma^{bc})_\alpha{}^\delta h_{abc}
+\frac{1}{48}(\gamma_{abcd})_\alpha{}^\delta g^{bcd}\,,
\end{equation}
where $h_{abc}$ and $g_{abc}$ are arbitrary 3-forms. The first Bianchi identity (\ref{eq:dim-1-I1}) then gives
\begin{equation}\te 
R_{\alpha\beta}{}^{cd}=\frac{i}{2}(\gamma_b)_{\alpha\beta}h^{bcd}+\frac{i}{24}\gamma^{cdefg}_{\alpha\beta}g_{efg}\,.
\end{equation}
The second Bianchi identity (\ref{eq:dim-1-I2}) now reads
\begin{equation}\te 
\nabla_{(\alpha}T_{\beta\gamma)}{}^\delta
+T_{(\alpha\beta}{}^\epsilon T_{\gamma)\epsilon}{}^\delta
-\frac{i}{4}\gamma^a_{(\alpha\beta}(\gamma^{bc})_{\gamma)}{}^\delta(h_{abc}-g_{abc})
=0\,.
\end{equation}
Contracting the indices $\gamma$ and $\delta$ and using the expression for the dimension $\ha$ torsion we find
\begin{equation}\te 
16\nabla_{(\alpha}\chi_{\beta)}-\gamma^a_{\alpha\beta}\gamma_a^{\gamma\delta}\nabla_\gamma\chi_\delta=0
\qquad\Rightarrow\qquad\nabla_{(\alpha}\chi_{\beta)}=\frac{i}{2}\gamma_{\alpha\beta}^aX_a\,,
\end{equation}
for some one-form superfield $X_a$. The remaining components of the Bianchi identity then give
\begin{equation}\te 
\nabla_\alpha\chi_\beta=
\chi_\alpha\chi_\beta
+\frac{i}{2}\gamma_{\alpha\beta}^aX_a
-\frac{i}{24}\gamma^{abc}_{\alpha\beta}(h_{abc}-g_{abc})\,, \la{a36}
%
%
%
\end{equation}
where we used the fact that $\chi_\alpha\chi_\beta=\frac{1}{96}\gamma^{abc}_{\alpha\beta}\,\chi\gamma_{abc}\chi$.

If we  finally  impose the 3-form $H=dB$ Bianchi identity 
and the kappa-symmetry constraints  on it in \rf{a1} 
we find, using the lower dimension constraints, that
\begin{equation}
T_{\alpha\beta}{}^eH_{cde}
+2T_{c(\alpha}{}^\gamma H_{\beta)\gamma d}
-2T_{d(\alpha}{}^\gamma H_{\beta)\gamma c}
=0\,,\la{a37}
\end{equation}
which implies that 
\begin{equation}
h_{abc}=H_{abc}\,,\qquad \qquad g_{abc}=0\,.
\end{equation}

In the {\bf type IIB} case the dimension 1 Bianchi identities are
\begin{align}
&2T_{c(\alpha i}{}^{\gamma k} T_{\beta j)\gamma k}{}^d=R_{\alpha i\beta jc}{}^d
\ , \\
&\te \nabla_{(\alpha i}T_{\beta j\gamma k)}{}^{\delta l}
+T_{(\alpha i\beta j}{}^{\epsilon m}T_{\gamma k)\epsilon m}{}^{\delta l}
+T_{(\alpha i\beta j}{}^aT_{|a|\gamma k)}{}^{\delta l}
=R_{(\alpha i\beta j\gamma k)}{}^{\delta l}\,.
\end{align}
The equations for $T_{a\beta i }{}^{\gamma i}$ and $R_{\alpha i \beta i c}{}^d$ with $i=1,2$ are the same as in the type I case analyzed above. This implies  (here primed and unprimed quantities are independent)
\begin{align}
&\te T_{a\alpha1}{}^{\delta1}=
\frac18(\gamma^{bc})_\alpha{}^\delta h_{abc}
+\frac{1}{48}(\gamma_{abcd})_\alpha{}^\delta g^{bcd}\,,\qquad
T_{a\alpha2}{}^{\delta2}=
\frac18(\gamma^{bc})_\alpha{}^\delta h'_{abc}
+\frac{1}{48}(\gamma_{abcd})_\alpha{}^\delta g'^{bcd}\,,
\\
&\te R_{\alpha1\beta1}{}^{cd}=\frac{i}{2}(\gamma_b)_{\alpha\beta}h^{bcd}+\frac{i}{24}\gamma^{cdefg}_{\alpha\beta}g_{efg}\,,
\qquad\ \ \ 
R_{\alpha2\beta2}{}^{cd}=\frac{i}{2}(\gamma_b)_{\alpha\beta}h'^{bcd}+\frac{i}{24}\gamma^{cdefg}_{\alpha\beta}g'_{efg}\,,
\\
&\te \nabla_{\alpha1}\chi_{\beta1}=\,
\chi_{\alpha1}\chi_{\beta1}
+\frac{i}{2}\gamma_{\alpha\beta}^a X_a 
-\frac{i}{24}\gamma^{abc}_{\alpha\beta}(h_{abc}-g_{abc})\,, \qquad\qquad  X_a \equiv \mathrm X_a+K_a\ , 
\la{a43}\\
&\te \nabla_{\alpha2}\chi_{\beta2}=\,
\chi_{\alpha2}\chi_{\beta2}
+\frac{i}{2}\gamma_{\alpha\beta}^a   X'_a 
-\frac{i}{24}\gamma^{abc}_{\alpha\beta}(h'_{abc}-g'_{abc})\,, \qquad\qquad X'_a \equiv \mathrm X_a-K_a\ .\la{a44}
\end{align}
Here instead of $X_a$ and $X'_a$   which appear as in type I case 
 we introduced the two independent 
 superfields   $\mathrm X_a $ and $ K_a$ for later convenience.

The remaining equations to solve are
\begin{align}
&\la{a45}R_{\alpha1\beta2c}{}^d
=
-iT_{c\alpha1}{}^{\gamma2}\gamma_{\beta\gamma}^d
-iT_{c\beta2}{}^{\gamma1}\gamma_{\alpha\gamma}^d\,,
\\  &\la{a46}
\nabla_{\gamma2}T_{\alpha1\beta1}{}^{\delta1}
-i\gamma_{\alpha\beta}^aT_{a\gamma2}{}^{\delta1}
=2R_{\gamma2(\alpha1\beta1)}{}^{\delta1}\,,
\\  &  \la{a47}
-i\gamma_{\beta\gamma}^aT_{a\alpha1}{}^{\delta1}=R_{\beta2\gamma2\alpha1}{}^{\delta1}\,,
\\  &\la{a48}
\gamma_{(\alpha\beta}{}^aT_{|a|\gamma2)}{}^{\delta1}=0\,,
\end{align}
together with the same equations with indices 1 and 2 interchanged. Eq. \rf{a47} implies
\begin{equation}
g_{abc}=g'_{abc}=0\,,\qquad \qquad h'_{abc}=-h_{abc}\,,
\end{equation}
while from \rf{a48}   we get
\begin{equation}\te 
T_{a\beta2}{}^{\gamma1}=\frac18(\gamma_a\mathcal S^{21})_\beta{}^\gamma\,,\qquad\qquad
T_{a\beta1}{}^{\gamma2}=\frac18(\gamma_a\mathcal S^{12})_\beta{}^\gamma\,,
\end{equation}
for some matrices $\mathcal S^{12}$ and $\mathcal S^{21}$. Eq. \rf{a45}  now implies
\begin{equation}\te
\mathcal S^{12}=-(\mathcal S^{21})^T
\ , \qquad \qquad 
R_{\alpha1\beta2}{}^{cd}=-\frac{i}{4}(\gamma^{[c}\mathcal S^{12}\gamma^{d]})_{\alpha\beta}\,.
\end{equation}
Finally, eq. \rf{a46}  gives
\begin{equation}\te 
\nabla_{\alpha2}\chi_\beta^1=-\frac{i}{16}(\gamma_a\mathcal S^{21}\gamma^a)_{\alpha\beta}\,.
\end{equation}
This completes the solution of the dimension 1 torsion Bianchi identities. 
The 3-form Bianchi identity just adds, as in the type I case,  the relation 
\begin{equation}
h_{abc}=H_{abc}\ .
\end{equation}

\subsection*{Dimension $3\ov 2$}

The {\bf type I} dimension $3\ov 2$ Bianchi identities are
\begin{align}
&\la{a54}-i\gamma^d_{\alpha\beta}T_{bc}{}^\beta=2R_{\alpha [bc]}{}^d
\ , \\
&\la{a55}\nabla_aT_{\beta\gamma}{}^\delta
-2\nabla_{(\beta}T_{|a|\gamma)}{}^\delta
+2T_{a(\beta}{}^\epsilon T_{\gamma)\epsilon}{}^\delta
-T_{\beta\gamma}{}^\epsilon T_{a\epsilon}{}^\delta
-i\gamma^e_{\beta\gamma}T_{ea}{}^\delta
=2R_{a(\beta\gamma)}{}^\delta\,.
\end{align}
Eq.\rf{a54}  gives the dimension $3\ov 2$ component of the curvature  as 
\begin{equation}\te
R_{\alpha bcd}=\frac{i}{2}(\gamma_b\psi_{cd})_\alpha-i(\gamma_{[c}\psi_{d]b})_\alpha\,,
\end{equation}
where $\psi_{ab}^\beta=T_{ab}{}^\beta$ is the gravitino field strength. Using this in \rf{a55} we get
\begin{align}&\te
\nabla_aT_{\beta\gamma}{}^\delta
-2\nabla_{(\beta}T_{|a|\gamma)}{}^\delta
+2T_{a(\beta}{}^\epsilon T_{\gamma)\epsilon}{}^\delta
-T_{\beta\gamma}{}^\epsilon T_{a\epsilon}{}^\delta
-\frac{i}{4}(\gamma^{cd})^\delta{}_{(\beta}(\gamma_a\psi_{cd})_{\gamma)}
\te
\nonumber\\ &\te
-\frac{i}{4}\gamma^c_{\beta\gamma}(\gamma_c\gamma^b\psi_{ba})^\delta
-\frac{i}{2}\gamma^b_{\beta\gamma}\psi_{ba}^\delta
+\frac{i}{2}\delta^\delta{}_{(\beta}(\gamma^b\psi_{ba})_{\gamma)}
=0\,.
\la{a57}\end{align}
Contracting the indices $\gamma$ and $\delta$ and using the lower dimension constraints gives
\begin{align}
&\te
i(\gamma^b\psi_{ab})_\alpha
={}
4\nabla_a\chi_\alpha
+(\gamma_a\gamma^b\nabla_b\chi)_\alpha
-\frac{1}{56}(\gamma_a\gamma^{bcd})_\alpha{}^\beta \nabla_\beta(h_{bcd}+\frac{11}{6}g_{bcd})
\nonumber\\
&\te {}
-\frac{1}{14}(\gamma^{bc})_\alpha{}^\beta \nabla_\beta(h_{abc}-\frac12g_{abc})
+\frac18(\gamma_a\gamma^{bcd}\chi)_\alpha(h_{bcd}+\frac{11}{6}g_{bcd})
+\frac12(\gamma^{bc}\chi)_\alpha(h_{abc}-\frac12g_{abc})\,. \la{a58}
\end{align}
Contracting \rf{a57} with $\gamma_e^{\beta\gamma}$  and using \rf{a58} 
we get, after some tedious algebra,
\begin{align}
&\te (\gamma^a)^{\alpha\beta}\nabla_\beta h_{abc}
=
-4(\gamma_{[b}\nabla_{c]}\chi)^\alpha
+\frac{5}{84}(\gamma_{bc}\gamma^{def})^{\alpha\beta} \nabla_\beta g_{def}
-\frac{3}{7}(\gamma_{[b}\gamma^{de})^{\alpha\beta}\nabla_\beta g_{c]de}
-\frac{1}{2}(\gamma^a)^{\alpha\beta}\nabla_\beta g_{abc}
\nonumber\\
&{}\te 
-\frac12(\gamma_{[b}\gamma^{de}\chi)^\alpha h_{c]de}
-\frac{13}{28}(\gamma_{bc}\gamma^{def}\chi)^\alpha g_{def}
+\frac{95}{28}(\gamma_{[b}\gamma^{de}\chi)^\alpha g_{c]de}
+4(\gamma^a\chi)^\alpha g_{abc}
+6i\psi_{bc}^\alpha\,.
\end{align}
Contracting \rf{a57}  with $(\gamma^{ef})_\delta{}^\gamma$ 
gives
\begin{align}&\te
\nabla_\alpha h_{abc}
=
3i(\gamma_{[a}\psi_{bc]})_\alpha
+\frac{1}{60}(\gamma_{abc}\gamma^{def})_\alpha{}^\beta \nabla_\beta g_{def}
-\frac{3}{20}(\gamma_{[ab}\gamma^{de})_\alpha{}^\beta \nabla_\beta g_{c]de}
-\frac{3}{10}(\gamma_{[a}\gamma^d)_\alpha{}^\beta \nabla_\beta g_{bc]d}
\nonumber\\
&{}\te
+\frac{1}{10}\nabla_\alpha g_{abc}
-\frac{2}{15}(\gamma_{abc}\gamma^{def}\chi)_\alpha g_{def}
+\frac{6}{5}(\gamma_{[ab}\gamma^{de}\chi)_\alpha g_{c]de}
+\frac{12}{5}(\gamma_{[a}\gamma^d\chi)_\alpha g_{bc]d}
-\frac45\chi_\alpha g_{abc}\,.\la{a60}
\end{align}
Using this in \rf{a57}  it  finally becomes 
\begin{equation}
\te -\frac{5}{3}(\gamma^{abcd})_{(\gamma}{}^\delta\langle\nabla_{\beta)}g^{bcd}+2\chi_{\beta)}g^{bcd}\rangle
-(\gamma^{cd})_{(\gamma}{}^\delta\langle\nabla_{\beta)}g^{acd}+2\chi_{\beta)}g^{acd}\rangle
%
=0\,,
\end{equation}
where we use the angle-brackets to denote the gamma-traceless part, e.g., 
\begin{equation}\te
\langle\nabla_\alpha g_{abc}\rangle=
\nabla_\alpha g_{abc}
+\frac{1}{21\cdot16}(\gamma_{abc}\gamma^{def})_\alpha{}^\beta \nabla_\beta g_{def}
-\frac{1}{14}(\gamma_{[ab}\gamma^{de})_\alpha{}^\beta \nabla_\beta g_{c]de}
-\frac12(\gamma_{[a}\gamma^d)_\alpha{}^\beta\nabla_\beta g_{bc]d}\,.\qquad
\end{equation}
This equation is easily shown to imply
\begin{equation}
\langle\nabla_\alpha g_{abc}+2\chi_\alpha g_{abc}\rangle=0\,.
\end{equation}
Using this in the expressions  \rf{a58} and \rf{a60}  they become
\begin{align}
&\te i(\gamma^b\psi_{ab})_\alpha
=
2\nabla_a\chi_\alpha
+\frac14(\gamma^{bc}\chi)_\alpha h_{abc}
+\frac{1}{84}(\gamma^{abcd})_\alpha{}^\beta \nabla_\beta g_{bcd}
-\frac{17}{168}(\gamma^{abcd}\chi)_\alpha g_{bcd}\,, \la{a64}\\
&\te \nabla_\alpha h_{abc}
=
3i(\gamma_{[a}\psi_{bc]})_\alpha
+\frac{11}{21\cdot32}(\gamma_{abc}\gamma^{def})_\alpha{}^\beta \nabla_\beta g_{def}
-\frac17(\gamma_{[ab}\gamma^{de})_\alpha{}^\beta \nabla_\beta g_{c]de} -\frac14(\gamma_{[a}\gamma^d)_\alpha{}^\beta \nabla_\beta g_{bc]d}
\nonumber\\
&{}\te
-\frac{15}{7\cdot16}(\gamma_{abc}\gamma^{def}\chi)_\alpha g_{def}
+\frac{17}{14}(\gamma_{[ab}\gamma^{de}\chi)_\alpha g_{c]de}
+\frac52(\gamma_{[a}\gamma^d\chi)_\alpha g_{bc]d}
-\chi_\alpha g_{abc}
\end{align}
This completes the solution of the torsion Bianchi identity.

It remains to analyze the consequence of the constraint \rf{a36} 
 on $\nabla_\alpha\chi_\beta$ found at dimension one. 
 To do this we take a symmetrized spinor derivative of this equation which gives 
\begin{align}\te
&2T_{\alpha\gamma}{}^D\nabla_D\chi_\beta
+2R_{\alpha\gamma\beta}{}^\delta\chi_\delta
+4\nabla_{(\alpha}\chi_{\gamma)}\chi_\beta\nonumber\\  &\te
-4\chi_{(\alpha}\nabla_{\gamma)}\chi_\beta
+2i\gamma_{\beta(\alpha}^a\nabla_{\gamma)}X_a
+\frac{i}{6}\gamma^{abc}_{\beta(\alpha}(\nabla_{\gamma)}h_{abc}-\nabla_{\gamma)}g_{abc})
=0\,.
\end{align}
Using the above expressions 
 we find the equation
\begin{align}
&\te 2\gamma_{\beta(\alpha}^a\nabla_{\gamma)}X_a
+\gamma^a_{\alpha\gamma}(\gamma_a\gamma^b\chi)_\beta X_b
-2\gamma^a_{\beta(\alpha}\nabla_a\chi_{\gamma)}
-\frac12\gamma^a_{\alpha\gamma}(\gamma_a\gamma^b\nabla_b\chi)_\beta
\nonumber\\
&{}\te 
+\frac{1}{96}\gamma^a_{\alpha\gamma}(\gamma_a\gamma^{def})_\beta{}^\delta\nabla_\delta g_{def}
+\frac{1}{48}\gamma^a_{\alpha\gamma}(\gamma_a\gamma^{bcd}\chi)_\beta h_{bcd}
-\frac{1}{4}\gamma^a_{\beta(\alpha}(\gamma^{bc}\chi)_{\gamma)} h_{abc}
-\frac{1}{32}\gamma^a_{\alpha\gamma}(\gamma_a\gamma^{bcd}\chi)_\beta g_{bcd}
\nonumber\\
&{}\te
-\frac{1}{8}\gamma^a_{\alpha\gamma}(\gamma^{de}\chi)_\beta g_{ade}
+\frac{5}{8}\gamma^a_{\beta(\alpha}(\gamma^{de}\chi)_{\gamma)} g_{ade}
+\frac{1}{6}\gamma^{abc}_{\beta(\alpha}\chi_{\gamma)} g^{abc}
+\frac{1}{48}\gamma^{abcde}_{\alpha\gamma}(\gamma_{de}\chi)_\beta g_{abc}
=0\,.\la{a67}
\end{align}
Contracting this with $\gamma_d^{\alpha\gamma}$ gives, after a bit of algebra,
\begin{align}
&\te \nabla_\alpha X_a
=
\frac12(\gamma^b\gamma_a\nabla_b\chi)_\alpha
+(\gamma_a\gamma^b\chi)_\alpha  X_b
+\frac{1}{96}(\gamma_a\gamma^{bcd})_\alpha{}^\beta\nabla_\beta g_{bcd}
+\frac{1}{48}(\gamma_a\gamma^{bcd}\chi)_\alpha h_{bcd}
\nonumber\\
&{}\te\qquad\qquad 
+\frac{1}{8}(\gamma^{bc}\chi)_\alpha h_{abc}
-\frac{7}{96}(\gamma_a\gamma^{bcd}\chi)_\alpha g_{bcd}
-\frac{1}{16}(\gamma^{bc}\chi)_\alpha g_{abc}\,.
\end{align}
It is not hard to show that this solves \rf{a67}.
This completes the solution of the torsion Bianchi identity in the type I case.

Imposing the Bianchi identity for the 3-form gives 
 no new constraints beyond what follows from the constraints 
 found  at dimension one, i.e. $h_{abc}=H_{abc}$ and $g_{abc}=0$.

In the {\bf  type IIB} case  the dimension $3\ov 2$ Bianchi identities are
\begin{align}
&\te T_{\alpha i\beta j}{}^dT_{bc}{}^{\beta j}=2R_{\alpha i[bc]}{}^d\ , \la{a69}
\\  &\te
\nabla_aT_{\beta i\gamma j}{}^{\delta k}
-2\nabla_{(\beta i}T_{|a|\gamma j)}{}^{\delta k}
+2T_{a(\beta i}{}^{\epsilon l} T_{\gamma j)\epsilon l}{}^{\delta k}\nonumber\\ &\te \qquad
-T_{\beta i\gamma j}{}^{\epsilon l} T_{a\epsilon l}{}^{\delta k}
-i\delta_{ij}\gamma^e_{\beta\gamma}T_{ea}{}^{\delta k}
=2R_{a(\beta i\gamma j)}{}^{\delta k}\,. \la{a70}
\end{align}
The first  gives  again  the dimension $3\ov 2$ component of the curvature as
\begin{equation}\te
R_{\alpha ibcd}=\frac{i}{2}(\gamma_b\psi_{cd})_{\alpha i}-i(\gamma_{[c}\psi_{d]b})_{\alpha i}\,.
\end{equation}
Eq. \rf{a70}  with $i=j=k$ is the same as in the type I case and the solution is therefore (note that
 in the type IIB case 
 $g_{abc}=0$ and $h_{abc}=H_{abc}$)
\begin{align}  &\te
\nabla_{\alpha i}H_{abc}=3i(\gamma_{[a}\sigma^3\psi_{bc]})_{\alpha i}
\ , \\  &\te
i(\gamma^b\psi_{ab})_{\alpha i}=2\nabla_a\chi_{\alpha i}+\frac14(\gamma^{bc}\sigma^3\chi)_{\alpha i}H_{abc}\,.\la{a73}
\end{align}
The remaining components of the Bianchi identity are
\begin{align} &\la{a74}
-2\nabla_{(\beta 1}T_{|a|\gamma 1)}{}^{\delta 2}
-T_{\beta1\gamma1}{}^{\epsilon1} T_{a\epsilon1}{}^{\delta2}
-i\gamma^e_{\beta\gamma}T_{ea}{}^{\delta 2}
=0
\ , \\  &\te\la{a75}
-2\nabla_{(\beta2}T_{|a|\gamma1)}{}^{\delta2}
+T_{a\gamma1}{}^{\epsilon2} T_{\beta2\epsilon2}{}^{\delta2}
+R_{\gamma1a\beta2}{}^{\delta2}
=0\ , 
\end{align}
and the same with indices 1 and 2 interchanged. Eq.\rf{a75} gives
\begin{align}
\nabla_{\alpha2}\mathcal S^{\beta1\gamma2}
=&
\delta_\alpha^\gamma\mathcal S^{\beta1\delta2}\chi^2_\delta
+\mathcal S^{\beta1\gamma2}\chi^2_\alpha
-\mathcal S^{\beta1\delta2}\gamma^a_{\alpha\delta}(\gamma_a\chi^2)^\gamma
-2i(\gamma^{ab})^\gamma{}_\alpha\psi_{ab}^{\beta1}\,,
\nonumber\\
\nabla_{\alpha1}\mathcal S^{\beta1\gamma2}
=&
\delta_\alpha^\beta\mathcal S^{\delta1\gamma2} \chi^1_\delta
+\mathcal S^{\beta1\gamma2}\chi^1_\alpha
-\mathcal S^{\delta1\gamma2}\gamma^a_{\alpha\delta}(\gamma_a\chi^1)^\beta
+2i(\gamma^{cd})^\beta{}_\alpha\psi_{cd}^{\gamma2}\,.
\end{align}
Eq.\rf{a74}  is then automatically satisfied.

 It remains to analyze the consequences of the dimension one conditions on $\nabla_{\alpha i}\chi_{\beta j}$ in \rf{a43}   and \rf{a44}.
  Applying  $\nabla_{\gamma k}$ and symmetrizing the derivatives we get, for $i=j=k$, the same condition as in the type I case  but now not for one $X_a$ but    two  vectors  $\XX_a\pm K_a$
\begin{align}&\te
\nabla_{\alpha 1}(\mathrm X_a+K_a)
=
\frac12(\gamma^b\gamma_a\nabla_b\chi)_{\alpha 1}
+(\gamma_a\gamma^b\chi)_{\alpha 1}(\mathrm X_b+K_b)\nonumber \\ &\qquad\qquad \qquad\qquad  \te
+\frac{1}{48}(\gamma_a\gamma^{bcd}\sigma^3\chi)_{\alpha 1} H_{bcd}
+\frac18(\gamma^{bc}\sigma^3\chi)_{\alpha 1} H_{abc}\,,
\\&\te
\nabla_{\alpha 2}(\mathrm X_a-K_a)
=
\frac12(\gamma^b\gamma_a\nabla_b\chi)_{\alpha 2}
+(\gamma_a\gamma^b\chi)_{\alpha 2}(\mathrm X_b-K_b) \nonumber \\ &\qquad \qquad \qquad \qquad\te
+\frac{1}{48}(\gamma_a\gamma^{bcd}\sigma^3\chi)_{\alpha 2} H_{bcd}
+\frac18(\gamma^{bc}\sigma^3\chi)_{\alpha 2} H_{abc}\,.
\end{align}
The remaining equations involve $\nabla_{(\gamma 1}\nabla_{\alpha 1)}\chi_{\beta 2}$ and $\nabla_{(\gamma 1}\nabla_{\alpha 2)}\chi_{\beta 1}$ giving 
\begin{align}  &\te
-i\gamma^a_{\alpha\gamma}\nabla_a\chi_{\beta 2}
+T_{\alpha1\gamma1}^{\delta1}\nabla_{\delta 1}\chi_{\beta 2}
+\frac14R_{\alpha 1\gamma 1}{}^{cd}(\gamma_{cd}\chi)_{\beta 2}
-\frac{i}{8}(\gamma_a\nabla_{(\gamma 1}\mathcal S^{12}\gamma^a)_{\alpha)\beta}
=0
\ , \\  &\te
\frac14R_{\gamma 1\alpha 2}{}^{cd}(\gamma_{cd}\chi)_{\beta 1}
-\frac{i}{16}(\gamma_a\nabla_{\gamma 1}\mathcal S^{21}\gamma^a)_{\alpha\beta}
+\nabla_{\alpha 2}\chi_{\gamma1}\chi_{\beta1}
-\chi_{\gamma1}\nabla_{\alpha 2}\chi_{\beta1}
\nonumber\\  &\te \qquad
{}+\frac{i}{2}\gamma_{\gamma\beta}^a\nabla_{\alpha 2}(\mathrm X_a+K_a)
-\frac{i}{24}\gamma^{abc}_{\gamma\beta}\nabla_{\alpha 2}H_{abc}
\,=0\ ,
%
\end{align}
and the same with indices $1$ and $2$ interchanged. 
Using the results derived so far it is easy to check that the first equation is automatically 
satisfied while the second determines the remaining spinor derivatives of  $\mathrm X_a\pm K_a$
\begin{align}
\nabla_{\alpha 2}(\mathrm X_a+K_a)
=&\te
\nabla_a\chi_{\alpha 2}
+\frac18(\gamma^{bc}\sigma^3\chi)_{\alpha 2} H_{abc}
+\frac18(\gamma^a\mathcal S\chi)_{\alpha 2}\,,
\\
\nabla_{\alpha 1}(\mathrm X_a-K_a)
=&\te
\nabla_a\chi_{\alpha 1}
+\frac18(\gamma^{bc}\sigma^3\chi)_{\alpha 1} H_{abc}
+\frac18(\gamma^a\mathcal S\chi)_{\alpha 1}\,.
\end{align}
This completes the solution of the dimension $3\ov 2$ torsion Bianchi identities. Imposing the Bianchi identity for the 3-form gives no new constraints.

\subsection*{Dimension 2}

The  {\bf type I}   Bianchi identities at dimension two read
\begin{equation}
R_{[abc]}{}^d=0
\ , \qquad \qquad 
\nabla_\alpha T_{bc}{}^\delta
+2\nabla_{[b}T_{c]\alpha}{}^\delta
+2T_{[b|\alpha|}{}^\beta T_{c]\beta}{}^\delta
+T_{bc}{}^\beta T_{\beta\alpha}{}^\delta
=R_{bc\alpha}{}^\delta\,.
\end{equation}
Using the above results for the lower dimensional components the latter becomes
\begin{align}
\nabla_\alpha\psi_{ab}^\delta
=&\te 
\frac14(\gamma^{cd})_\alpha{}^\delta
\big[
R_{ab}{}^{cd}
-\nabla_{[a}h_{b]cd}
+\frac12h_{ac}{}^eh_{bde}
-\frac18g_{ac}{}^eg_{bde}
+\frac18\eta_{c[a}g_{b]ef}g_d{}^{ef}
-\frac{1}{48}\eta_{ac}\eta_{bd}g_{efg}g^{efg}
\big]
\nonumber\\
&{}\te
+\frac{1}{48}(\gamma^{cdef})_\alpha{}^\delta\left[h_{abc}g_{def}-2\eta_{c[a}\nabla_{b]}g_{def}+3\eta_{c[a}h_{b]d}{}^gg_{efg}\right]
-\frac{1}{192}(\gamma^{cdefgh})_\alpha{}^\delta \eta_{c[a}g_{b]de}g_{fgh}
\nonumber\\
&{}\te
+\frac{1}{128}(\gamma_{abcdef})_\alpha{}^\delta g^{cd}{}_gg^{efg}
-\psi_{ab}{}^\delta\chi_\alpha
-\delta^\delta_\alpha\psi_{ab}\chi
+(\gamma^c\psi_{ab})_\alpha(\gamma_c\chi)^\delta\,.
\end{align}
Multiplying this with $\gamma^b_{\beta\delta}$ and using the dimension $3\ov 2$ constraint 
on the gamma-trace of $\psi_{ab}$ in \rf{a64}
as well as the other lower dimension constraints gives some of the equations of motion. 
Let us also use 
  the Bianchi identitiy for  the 3-form $H$ which,  as we have seen,  lead 
  to $g_{abc}=0$, $h_{abc}=H_{abc}$. 
  We  then obtain the equations of motion (\ref{eq:H-R-bianchi}) and (\ref{eq:einstein-eq-I}). The final equation of motion comes from evaluating $\nabla_{(\alpha}\nabla_{\beta)}X_a$ using the 
  consequences of the dimension $3\ov 2$ constraint. Setting $g_{abc}=0$ this gives the equation
  (\ref{eq:div-X-I})  for the divergence of $X_a$.


In the {\bf type IIB } case  the dimension 2 Bianchi identities are
\begin{equation}
R_{[abc]}{}^d=0
\ , \qquad \qquad 
\nabla_{\alpha i}T_{bc}{}^{\delta j}
+2\nabla_{[b}T_{c]\alpha i}{}^{\delta j}
+2T_{[b|\alpha i|}{}^{\beta k}T_{c]\beta k}{}^{\delta j}
+T_{bc}{}^{\beta k}T_{\beta k\alpha i}{}^{\delta j}
=R_{bc\alpha i}{}^{\delta j}\,.
\end{equation}
The latter gives
\begin{align}
\nabla_{\alpha i}\psi_{ab}^{\delta j}
=&\te
-\frac14\sigma^3_{ij}(\gamma^{cd})_\alpha{}^\delta\nabla_{[a}H_{b]cd}
+\frac14(\gamma_{[a}\nabla_{b]}\mathcal S)_{\alpha i}{}^{\delta j}
+\frac{1}{8}\delta_{ij}(\gamma^{cd})_\alpha{}^\delta H_{ace}H_{bd}{}^e
-\frac14R_{ab}{}^{cd}\delta_{ij}(\gamma_{cd})^\delta{}_\alpha
\nonumber\\
&{}\te
+\frac{1}{32}(\gamma^{cd}\sigma^3\gamma_{[a}\mathcal S)_{\alpha i}{}^{\delta j}H_{b]cd}
-\frac{1}{32}(\gamma_{[a}\mathcal S\gamma^{cd}\sigma^3)_{\alpha i}{}^{\delta j}H_{b]cd}
-\frac{1}{32}(\gamma_{[a}\mathcal S\gamma_{b]}\mathcal S)_{\alpha i}{}^{\delta j}
\nonumber\\
&{}\te
-\frac12\delta_{ij}\delta^\delta_\alpha\psi_{ab}\chi
-\frac12\sigma^3_{ij}\delta^\delta_\alpha\psi_{ab}\sigma^3\chi
-\frac12\psi_{ab}^{\delta j}\chi_{\alpha i}
-\frac12(\sigma^3\psi_{ab})^{\delta j}(\sigma^3\chi)_{\alpha i}
\nonumber\\
&{}\te
+\frac12(\gamma^c\psi_{ab})_{\alpha i}(\gamma_c\chi)^{\delta j}
+\frac12(\gamma^c\sigma^3\psi_{ab})_{\alpha i}(\gamma_c\sigma^3\chi)^{\delta j}\,.
\label{eq:nabla-psi-IIB}
\end{align}
Multiplying this with $\gamma^a_{\beta\delta}$ and using the dimension $3\ov 2$ constraint \rf{a73}
on $\gamma^a\psi_{ab}$ as well as the lower dimension constraints we get 
\begin{align}  &\te
\delta_{ij}\gamma^a_{\alpha\beta}\nabla_b\mathrm X_a
+\frac12\delta_{ij}\gamma^c_{\alpha\beta}K^aH_{abc}
-\frac14\sigma^3_{ij}\gamma^c_{\alpha\beta}(\nabla^aH_{abc}-2{\mathrm X}^aH_{abc})
\nonumber\\  &\te
+\sigma^3_{ij}\gamma^a_{\alpha\beta}\nabla_bK_a
+\frac16\sigma^3_{ij}\gamma^{cde}_{\alpha\beta}\nabla_{[b}H_{cde]}
-\frac18(\gamma_b\nabla_a\mathcal S\gamma^a)_{\alpha i\beta j}
+\frac18(\gamma_b\mathcal S\gamma^a)_{\alpha i\beta j}\mathrm X_a
\nonumber\\  &\te
+\frac18(\gamma_b\mathcal S\sigma^3\gamma^a)_{\alpha i\beta j}K_a
-\frac18\delta_{ij}\gamma^a_{\alpha\beta}H_{acd}H_b{}^{cd}
-\frac14R_{ab}{}^{cd}\delta_{ij}(\gamma^a\gamma_{cd})_{\beta\alpha}
\nonumber\\  &\te
+\frac{1}{192}(\gamma_b\mathcal S\gamma^{cde}\sigma^3)_{\alpha i\beta j}H_{cde}
+\frac{1}{16}(\gamma^c\mathcal S\gamma^d\sigma^3)_{\alpha i\beta j}H_{bcd}
-\frac{1}{64}(\gamma^{cd}\sigma^3\gamma_b\mathcal S\gamma^a)_{\alpha i\beta j}H_{acd}
\nonumber\\  &\te
-\frac{1}{64}(\gamma_a\mathcal S\gamma_b\mathcal S\gamma^a)_{\alpha i\beta j}
-\frac{i}{8}(\gamma_b\mathcal S\chi)_{\alpha i}\chi_{\beta j}
-\frac{i}{8}(\gamma_b\mathcal S\sigma^3\chi)_{\alpha i}(\sigma^3\chi)_{\beta j}
-\frac12\delta_{ij}\gamma^a_{\alpha\beta}\psi_{ab}\chi
\nonumber\\  &\te
-\frac12\sigma^3_{ij}\gamma^a_{\alpha\beta}\psi_{ab}\sigma^3\chi
-\frac14(\gamma_b\psi_{cd})_{\alpha i}(\gamma^{cd}\chi)_{\beta j}
+\frac14(\gamma_b\sigma^3\psi_{cd})_{\alpha i}(\gamma^{cd}\sigma^3\chi)_{\beta j}
=0\,,
\end{align}
which implies the  equations \rf{eq:dH} and (\ref{eq:einstein-eq-IIB})--(\ref{eq:RR-eom}).

In addition, we have the consistency conditions that come from applying  two symmetrized spinor derivatives to a dimension 1 superfield and using the dimension $3\ov 2$  constraints. Doing this on the equations for the spinor derivative of $H_{abc}$ and $\mathcal S$ gives nothing new,  but from the equations for the derivative of $\mathrm X_a$ and $K_a$ we get
\begin{align}
&\te -\frac{i}{4}(\gamma_a(1+\sigma^3))_{\alpha i\beta j}
\big[
\nabla^b(\mathrm X_b+K_b)
-2({\mathrm X}^b+K^b)(\mathrm X_b+K_b)
+\frac{1}{12}H^{bcd}H_{bcd}
-\frac{1}{256}\mathrm{Tr}(\mathcal S\gamma^b\mathcal S\gamma_b)
\nonumber\\ &\te \qquad
+i\chi\gamma^b(1+\sigma^3)\nabla_b\chi
-\frac{i}{24}\chi\gamma^{bcd}(1+\sigma^3)\chi\,H_{bcd}
\big]
+\ldots
=0\ , 
\\
&\te -\frac{i}{4}(\gamma_a(1-\sigma^3))_{\alpha i\beta j}
\big[
\nabla^b(\mathrm X_b-K_b)
-2({\mathrm X}^b-K^b)({\mathrm X}_b-K_b)
+\frac{1}{12}H^{bcd}H_{bcd}
-\frac{1}{256}\mathrm{Tr}(\mathcal S\gamma^b\mathcal S\gamma_b)
\nonumber\\  &\te \qquad
+i\chi\gamma^b(1-\sigma^3)\nabla_b\chi
+\frac{i}{24}\chi\gamma^{bcd}(1-\sigma^3)\chi\,H_{bcd}
\big]
+\ldots
=0\,,
\end{align}
where the ellipsis denotes terms that vanish 
upon use of  the other equations of motion. These give the remaining equations of motion for the bosonic superfields \rf{eq:dh-etc} and \rf{eq:div-K}.
The Bianchi identity for the 3-form gives no new constraints.

\subsection*{Dimension $5\ov 2$}

The highest component of the {\bf type I}  torsion Bianchi identity reads
\begin{equation}
\nabla_{[a}T_{bc]}{}^\alpha-T_{[ab}{}^\beta T_{c]\beta}{}^\alpha=0\,.
\end{equation}
It gives the Bianchi identity for the gravitino field strength
\begin{equation}\te
\nabla_{[a}\psi_{bc]}^\alpha+\frac18(\gamma^{de}\psi_{[ab})^\alpha\,h_{c]de}-\frac{1}{48}(\gamma_{def[a}\psi_{bc]})^\alpha\,g^{def}=0\,.
\end{equation}
The condition coming from the symmetrized spinor derivative of $\psi_{ab}$ using the dimension $3\ov 2$ constraint on $\nabla_\alpha\psi_{ab}$ gives an equation for $\nabla_\alpha R_{ab}{}^{cd}$.
The latter is more easily obtained from the curvature Bianchi identity (\ref{eq:curvature-bianchi}) and reads
\begin{align}
\nabla_\alpha R_{ab}{}^{cd}
=&\te
-i(\gamma_{[a}\nabla_{b]}\psi^{cd})_\alpha
-i(\gamma^{[c}\nabla^{d]}\psi_{ab})_\alpha
-\frac{i}{8}(\gamma^{ef}\gamma_{[a}\psi^{cd})_\alpha h_{b]ef}
-\frac{i}{8}(\gamma_{ef}\gamma^{[c}\psi_{ab})_\alpha h^{d]ef}
\nonumber\\
&{}\te
+i(\gamma_e\psi^{[c}{}_{[a})_\alpha h_{b]}{}^{d]e}
-\frac{i}{48}(\gamma_{efg[a}\gamma_{b]}\psi^{cd})_\alpha g^{efg}
-\frac{i}{16}(\gamma^{cdefg}\psi_{ab})_\alpha g_{efg}
\nonumber\\
&{}\te
-\frac{i}{16}(\gamma^{ef[c}\psi_{ab})_\alpha g^{d]}{}_{ef}
-\frac{i}{4}(\gamma_{ef[a}\psi_{b]}{}^{[c})_\alpha g^{d]ef}
+\frac{i}{12}(\gamma^{efg}\psi_{[a}{}^{[c})_\alpha\delta^{d]}_{b]} g_{efg}\,.
\end{align}
Similarly, in  the {\bf type IIB} case we find 
\begin{align}
&\te \nabla_{[a}\psi_{bc]}^{\alpha i}
+\frac18(\gamma^{de}\sigma^3\psi_{[ab})^{\alpha i}\,H_{c]de}
+\frac18(\mathcal S\gamma_{[a}\psi_{bc]})^{\alpha i}
=0\,, \\
&\te\nabla_{\alpha i}R_{ab}{}^{cd}
=
-i(\gamma_{[a}\nabla_{b]}\psi^{cd})_{\alpha i}
-i(\gamma^{[c}\nabla^{d]}\psi_{ab})_{\alpha i}
-\frac{i}{8}(\gamma^{ef}\gamma_{[a}\sigma^3\psi^{cd})_{\alpha i}H_{b]ef}
-\frac{i}{8}(\gamma_{ef}\gamma^{[c}\sigma^3\psi_{ab})_{\alpha i}H^{d]ef}
\nonumber\\
&{}\te\qquad\qquad\quad
+i(\gamma_e\sigma^3\psi^{[c}{}_{[a})_{\alpha i}H_{b]}{}^{d]e}
+\frac{i}{8}(\gamma_{[a}\mathcal S\gamma_{b]}\psi^{cd})_{\alpha i}
+\frac{i}{8}(\gamma^{[c}\mathcal S\gamma^{d]}\psi_{ab})_{\alpha i}
+\frac{i}{4}(\gamma_{[a}\mathcal S\gamma^{[c}\psi_{b]}{}^{d]})_{\alpha i}
\nonumber\\
&{}\te\qquad\qquad\quad
-\frac{i}{4}(\gamma^{[c}\mathcal S\gamma_{[a}\psi_{b]}{}^{d]})_{\alpha i}\,.
\end{align}

\end{document}